\documentclass[iop]{emulateapj}
\pdfoutput=1
\usepackage{natbib}
\usepackage{graphicx}

\received{receipt date}
\revised{revision date}
\accepted{acceptance date}

\shorttitle{The disk-outflow system in S255IR}
\shortauthors{Zinchenko, Liu, Su, et al.}
\def\htwo{H\,{\scriptsize\sc{II}}}
\def\Fetwo{Fe\,{\scriptsize\sc{II}}}

\begin{document}

\title{The disk-outflow system in the S255IR area of high mass star formation}
\author{I. Zinchenko\altaffilmark{1,2}, S.-Y. Liu\altaffilmark{3}, Y.-N. Su\altaffilmark{3}, S. V. Salii\altaffilmark{4}, A. M. Sobolev\altaffilmark{4}, P. Zemlyanukha\altaffilmark{1,2}, H. Beuther\altaffilmark{5},\\ D. K. Ojha\altaffilmark{6}, M. R. Samal\altaffilmark{7}, Y. Wang\altaffilmark{8,9}}
\altaffiltext{1}{Institute of Applied Physics of the Russian Academy of Sciences, 46 Ulyanov st., Nizhny Novgorod 603950, Russia}
\altaffiltext{2}{Lobachevsky State University of Nizhni Novgorod, 23 Gagarin av., Nizhny Novgorod 603950, Russia}
\email{zin@appl.sci-nnov.ru}
\altaffiltext{3}{Institute of Astronomy and Astrophysics, Academia Sinica.
P.O. Box 23-141, Taipei 10617, Taiwan, R.O.C.}
\altaffiltext{4}{Ural Federal University, Ekaterinburg, Russia}
\altaffiltext{5}{Max-Planck-Institut f\"ur Astronomie, Heidelberg, Germany}
\altaffiltext{6}{Infrared Astronomy Group, Department of Astronomy and Astrophysics, Tata Institute  of  Fundamental Research, Homi Bhabha Road, Colaba,  Mumbai (Bombay) -- 400 005, India}
\altaffiltext{7}{Laboratoire d'Astrophysique de Marseille (UMR 6110 CNRS \& Universit\'e de Provence), 38 rue F. Joliot-Curie, 13388 Marseille Cedex 13, France}
\altaffiltext{8}{Department of Astronomy, University of Geneva, Switzerland}
\altaffiltext{9}{Purple Mountain Observatory, CAS, China}

\begin{abstract}
We report the results of our observations of the S255IR area with the SMA at 1.3~mm in the very extended configuration and at 0.8~mm in the compact configuration as well as with the IRAM-30m at 0.8~mm. The best achieved angular resolution is about 0.4 arcsec. The dust continuum emission and several tens of molecular spectral lines are observed. The majority of the lines is detected only towards the S255IR-SMA1 clump, which represents a rotating structure (probably disk) around the young massive star. The achieved angular resolution is still insufficient for conclusions about Keplerian or non-Keplerian character of the rotation. The temperature of the molecular gas reaches 130--180~K. The size of the clump is about 500~AU. The clump is strongly fragmented as follows from the low beam filling factor. The mass of the hot gas is significantly lower than the mass of the central star. A strong DCN emission near the center of the hot core most probably indicates a presence of a relatively cold ($\lesssim 80$~K) and rather massive clump there. High velocity emission is observed in the CO line as well as in lines of high density tracers HCN, HCO$^+$, CS and other molecules. The outflow morphology obtained from combination of the SMA and IRAM-30m data is significantly different from that derived from the SMA data alone. The CO emission detected with the SMA traces only one boundary of the outflow. The outflow is most probably driven by jet bow shocks created by episodic ejections from the center. We detected a dense high velocity clump associated apparently with one of the bow shocks. 
The outflow strongly affects the chemical composition of the surrounding medium. 
\end{abstract}

\section{Introduction}
{The formation of high mass stars (more massive than 8--10 M$_\odot$) is still poorly understood. Studies of this process are currently a ``hot topic" of astrophysical research.}
A major unsolved problem of high mass star formation is the characterization of accretion disks around young high-mass protostars. To date, most of evidence for the existence of such disks has been indirect \citep[e.g.][]{Beuther09}.  If present, these disks undoubtedly play a crucial role in the star formation process.  Nevertheless, the very existence of these disks, much less their physical properities, is not well-established.  The importance of understanding these disks cannot be understated.  For example, because of the larger masses involved, it is quite possible that high-mass disks are self-gravitating, unlike their Keplerian low-mass counterparts, implying that there may be significant dynamical differences between the high-mass and low-mass star formation processes.

S255IR is a part of the well known massive star forming complex located between the evolved Sharpless \htwo\ regions S255 and S257. Several authors estimated the photometric distance to the complex at about 2.5~kpc \citep{Russeil07,Chavarria08,Ojha11}. However, \citet{Rygl10} report a distance of 1.6~kpc based on trigonometric parallax measurements of methanol masers. This seems to be the most accurate distance estimate and we adopt it here, too. 
{The previous studies have shown that S255IR contains a cluster of early-B-type stars \citep{Howard97,Itoh01}, several compact \htwo\ regions \citep{Snell86}, and a number of
H$_2$ emission features \citep{Miralles97}. 
The total mass of the S255IR core has been estimated from single-dish observations at $M \sim 300-400$~M$_{\odot}$ \citep{Zin09,Wang11}. The estimate of the luminosity has been $L \sim 2\times 10^4$~L$_{\odot}$ \citep{Wang11}.}

 {The interferometric observations by \citet{Wang11} revealed 3 continuum clumps and a high-velocity collimated outflow in the S255IR area.}
Our Paper~I \citep{Zin12} was devoted to the general structure and kinematics of the complex based on the SMA observations at 1.3 and 1.1~mm ( {the synthesized beam sizes were approximately $3\farcs8 \times 3\farcs0$ at 1.3~mm and 
$2\farcs9 \times 2\farcs6$ at 1.1~mm}) as well as VLA ammonia observations and GMRT low frequency continuum data. Our results as well as data by \citet{Wang11} indicate in particular a presence of a hot rotating core and a spectacular outflow in the S255IR region. However the angular resolution was insufficient for a more detailed investigation of this core. In addition, a reliable evaluation of physical and chemical properties of this system required observations of additional molecular transitions. Therefore we performed observations of the S255IR area with the SMA at a much higher angular resolution and at higher frequencies. 

Here we present observational data for S255IR obtained at 1.3 mm with the SMA in the very extended configuration and at 0.8 mm in the compact configuration. In addition we observed this area with the IRAM 30m radio telescope in order to obtain short-spacing data complementing the SMA results. We mainly discuss properties of the dense cores and high velocity outflows observed in this area. 

\section{Observations and data reduction}

\subsection{SMA}
The S255~IR area was observed with the Submillimeter Array (SMA) in its compact
configuration on 2010 December 14th at 350~GHz.
A three field mosaic was obtained with the following phase centers:
06$^h$12$^m$53$^s$.800, 17$^\circ$59$'$22$''$.1,  06$^h$12$^m$54$^s$.8876, 17$^\circ$59$'$29$''$.3353 and 06$^h$12$^m$52$^s$.7124, 17$^\circ$59$'$14$''$.8647. The primary HPBW of the SMA antennas is 36$^{\prime\prime}$ at these frequencies.
Typical system temperatures on source were between 200~K and 400~K.
The resulting $uv$ coverage ranges from 10~k$\lambda$ to 90~k$\lambda$ (8 antennas in the array).
3C454.3 and Uranus were used as the bandpass calibrators 
and 0532+075, 0750+125 and 0530+135 were used as the complex phase and amplitude gain calibrators.
A total of 8~GHz (342$-$346~GHz in the LSB and 354$-$358~GHz in the USB)
was observed with the SMA bandwidth doubling correlator configuration.
The spectral resolution was 0.8125~MHz.

In addition on 2011 January 07th S255IR was observed in the very extended configuration at 225~GHz. A single field was observed with the same phase center as for the central field at 350~GHz. The primary HPBW of the SMA antennas is 55$^{\prime\prime}$ at these frequencies. Typical system temperatures on source were between 90~K and 200~K.
The resulting $uv$ coverage ranges from 50~k$\lambda$ to 350~k$\lambda$ (7 antennas in the array). 3C454.3 and 3C279 were used as the bandpass calibrators 
and 0750+125 and 0530+135 were used as the complex phase and amplitude gain calibrators.
A total of 8~GHz (217.0$-$220.8~GHz in the LSB and 229.1$-$232.95~GHz in the USB)
was observed with the SMA bandwidth doubling correlator configuration.
The spectral resolution was 0.406~MHz and 1.625~MHz ( {given the limited correlator capability, we were not able to have the high resolution across the whole band; so, we assigned different resolutions for different tracers}). These data were combined with our previous data obtained at the same frequencies in the compact configuration (Paper~I) which enables an investigation of the source structure in a wider range of spatial scales.

The gain calibrator flux scale, calibrated against Uranus at 350~GHz and Callisto at 225~GHz, was found to be consistent within 5\% with the SMA calibrator database
and estimated to be accurate within 20\%.

The data calibration was carried out with the IDL superset MIR \citep{Scoville93},
and subsequent imaging and analysis were done in MIRIAD \citep{Sault95}.
With robust weighting for the continuum and line data,
the synthesized beam sizes are approximately $2\farcs2 \times 1\farcs9$ at 350~GHz and 
$0\farcs5 \times 0\farcs4$ at 225~GHz. The rms noise is approximately 7~mJy\,beam$^{-1}$ and 1~mJy\,beam$^{-1}$ in the continuum images at 350~GHz and 225~GHz, respectively, and 100~mJy\,beam$^{-1}$ and 20~mJy\,beam$^{-1}$ in the spectral cubes at these frequencies at 2~km\,s$^{-1}$ resolution.


\subsection{IRAM 30m radio telescope}
Single-dish observations of several molecular lines at the 30m IRAM radio telescope were performed in October 2012 (N$_2$H$^+$ $ J=3-2 $ at 279.5~GHz), December 2012 (SiO $J=5-4 $ at 217.1~GHz) and January 2014 (CO $ J=3-2 $ at 345.8~GHz and CS $ J=7-6 $ at 342.9~GHz). The antenna HPBW was $9\farcs3$, $11\farcs9$ and $7\farcs5$ at these frequencies, respectively. The observations were performed in the OTF mode with position switching using the HERA receiver at 217.1~GHz and EMIR receiver at higher frequencies. The central position of the maps was the same as for the primary field in the SMA observations. The reference position was selected at $-$500$''$ in right ascension from the central position. Apparently a weak CO(3--2) emission is present at the reference position resulting in a weak negative feature in the CO spectra at $ V_\mathrm{LSR}\approx 24 $~km\,s$^{-1}$ (the bulk of the line emission in this area is observed at $V_\mathrm{LSR}\sim 4-10$~km\,s$^{-1}$). This feature  {does not affect significantly the observed spectra}. 
The system temperature was $ \sim 200 $~K for the N$_2$H$^+$ observations, $ \sim 300 $~K for the SiO observations and $ \sim 600 $~K for the CO and CS observations. Pointing was checked regularly on nearby strong sources and pointing errors were within a few arcseconds. The antenna temperature calibration was made by the standard chopper-wheel method.  

The map size was approximately $2\farcm0 \times 2\farcm5$ for the N$_2$H$^+$ and SiO observations, covering both S255IR and S255N regions, and approximately $1\farcm5 \times 1\farcm0$ for the CO and CS observations, covering only the S255IR area. Here we discuss only the data relevant to S255IR. The data on S255N are postponed for further publications.

The data reduction was performed with the GILDAS package (http://www.iram.fr/IRAMFR/GILDAS). Then the single-dish data were combined with the SMA data using the MIRIAD procedures as described by e.g. \citet{Wang11}. The conversion to the flux density scale was made using the conversion factors presented on the IRAM 30m telescope website. 

\section{Observational results and data analysis} \label{sec:results}
With the SMA we detected several tens of spectral lines in both 350~GHz and 225~GHz bands. A list of these lines, including their frequencies and energy of lower levels, is given in Tables~\ref{table:lines-vext}, \ref{table:lines-comp}. The spectral line parameters are taken from the JPL \citep{Pickett98} and CDMS \citep{Mueller01,Mueller05} catalogs.

We present the results in the form of maps as well as spectra and line parameters at selected positions. For continuum observations we give positions, flux densities, and size estimates of the continuum sources. 

\begin{deluxetable}{lcrrc}
\tablecaption{List of molecular transitions observed at the SMA in the very extended configuration in S255IR. \label{table:lines-vext}}
\tablehead{\colhead{Molecule} &\colhead{Transition} &\colhead{Frequency} &\colhead{$E_l$} & \\ \colhead{ } &\colhead{ } &\colhead{(GHz)} &\colhead{(K)} &}
\tablecolumns{5}
\startdata
$ ^{12} $CO	&2--1	&230.538000		&5.532	& \\
$ ^{13} $CO	&2--1	&220.398684		&5.289	& \\
CH$ _{3} $OH	&$6_{1} - 7_{2}$ A$^-$	&217.299202	&363.496 \\
&$15_{6} - 16_{5}$	A$^-$	&217.642677		&735.160	&\\
&$15_{6} - 16_{5}$	A$^+$	&217.642678	&735.160	&\\
&$20_{1} - 20_{0}$	E	&217.886504		&497.919	&\\
&$4_{2} - 3_{1}$	E	&218.440063		&34.976	&\\
&$8_{0} - 7_{1}$	E	&220.078561		&86.051	&\\
&$15_{4} - 16_{3}$ E	&229.589056		&363.420&\\
&$8_{-1} - 7_{0}$	E	&229.758756		&78.076	&\\
&$19_{5} - 20_{4}$	A$^+$	&229.864121		&567.565	&\\
&$19_{5} - 20_{4}$	A$^-$	&229.939095	&567.561	&\\
&$3_{-2} - 4_{-1}$	E	&230.027047		&28.788	&\\
&$22_{2} - 21_{-3}$	E	&230.292196		&598.499	&\\
&$10_{2} - 9_{3}$ A$^-$	&231.281110		&154.248&\\
&$10_{2} - 9_{3}$ A$^+$	&232.418521		&154.248\\
&$18_{3} - 17_{4}$ A$^+$	&232.783446		&435.360\\

$^{13}$CS	&5--4	&231.220686		&22.194	&\\
DCN		&3--2	&217.238530		&10.425	&\\
HNCO	&$ 10_{0,10}-9_{0,9} $	&219.798282		&47.471	&\\
		&$ 10_{1,9}-9_{1,8} $	&220.584762		&90.916	&\\
		&$ 10_{1,10}-9_{1,9} $	&218.981170		&90.569	\\
		&$ 10_{2,9}-9_{2,8} $	&219.733850		&217.739	&\\
		&$ 10_{2,8}-9_{2,7} $	&219.737193		&217.739	&\\
		&$ 10_{3,8}-9_{3,7} $	&219.656770		&422.417	&\\
		&$ 10_{3,7}-9_{3,6} $	&219.656771		&422.417	&\\
HC$_3$N	&24--23	&218.324711		&120.504	&\\
SO		&$6_5 -        5_4$	&219.949433		&24.429	&\\
H$_2$CO	&$3_{0,3} -    2_{0,2}$	&218.222195		&10.483	&\\
		&$3_{2,2} -    2_{2,1}$	&218.475642		&57.608	&\\
		&$3_{2,1} -    2_{2,0}$	&218.760071		&57.613	&\\
OCS		&18--17	&218.903357		&89.304 &\\
		&19--18	&231.060983		&99.810	&\\
HCOOH &$ 10_{0,10}-9_{0,9} $	&220.038072 &48.061 &\\
       &$ 10_{1,9}-9_{1,8} $	&231.505705 &53.355\\
CH$_3$CN &$12_{6} - 11_{6}$ &220.594423  &315.313 \\
&$12_{5} - 11_{5}$ &220.641084   &236.810 \\
&$12_{4} - 11_{4}$ &220.679287   &172.555 \\
&$12_{3} - 11_{3}$ &220.709017   &122.565 \\
&$12_{2} - 11_{2}$ &220.730261   &86.849 \\
&$12_{1} - 11_{1}$ &220.743011   &65.416 \\
&$12_{0} - 11_{0}$ &220.747261   &58.272 

\enddata

\end{deluxetable}

\begin{deluxetable}{lcrrc}
\tablecaption{List of molecular transitions observed at the SMA in the compact configuration in S255IR. \label{table:lines-comp}}
\tablehead{\colhead{Molecule} &\colhead{Transition} &\colhead{Frequency} &\colhead{$E_l$} & \\ \colhead{ } &\colhead{ } &\colhead{(GHz)} &\colhead{(K)} &}
\tablecolumns{5}
\startdata

$^{12}$CO &3--2	&345.795990	&16.596	& \\
CS 	&7--6	&342.882850 	&49.372	& \\
HCO$^+$	&$4 -    3$	&356.734242 	&25.682	&\\
CH$ _{3} $OH &$13_{1} - 13_{0}$	A$^{-+}$	&342.729796		&211.024	&\\
&$13_{-1} - 14_{-2}$ E &343.599019   &607.550&\\
&$18_{2} - 17_{3}$	E	&344.109039		&402.884	&\\
&$19_{1} - 18_{2}$	A$^+$	&344.443433		&434.697	&\\
&$16_{1} - 15_{2}$	A$^-$	&345.903916		&316.049	&\\
&$18_{-3} - 17_{-4}$	E	&345.919260		&442.829	&\\
&$13_{0} - 12_{1}$	A$^+$	&355.602945		&193.959	&\\
HCN	&4--3	&354.505473		&25.251	&\\
	&4--3 {(0,1$^{1c}$,0)}	&354.460435		&1049.892	&\\
	&4--3 {(0,1$^{1d}$,0)}	&356.255568		&1050.021	&\\
H$^{13}$CN	&4--3	&345.339769		&24.861	&\\
HC$^{15}$N	&4--3	&344.200109		&24.779	&\\
HC$_3$N	&38--37	&345.609016		&306.905	&\\
&39--38	&354.697456		&323.492	&\\
SO		&$8_8 -        7_7$	&344.310792		&70.957	&\\
$^{33}$SO &$8_{9,8} -  7_{8,7}$	& 343.086102 &61.564 \\
&$8_{9,9} -  7_{8,8}$	& 343.087298 &61.566 \\
SO$_2$		&$13_{2,12} -        12_{1,11}$	&345.338538		&76.410	&\\
&$12_{4,8} -        12_{3,9}$	&355.045517	&93.960	&\\
&$13_{4,10} -        13_{3,11}$	&357.165390 	&105.823	&\\
&$15_{4,12} -        15_{3,13}$	&357.241193		&132.537	&\\
&$11_{4,8} -        11_{3,9}$	&357.387580		&82.800	&\\
&$8_{4,4} -        8_{3,5}$	&357.581449		&55.202	&\\
&$9_{4,6} -        9_{3,7}$	&357.671821		&15.992	&\\
&$7_{4,4} -        7_{3,5}$	&357.892442 	&47.835	&\\
&$6_{4,2} -        6_{3,3}$	&357.925848 	&41.402	&\\
&$17_{4,14} -        17_{3,15}$	&357.962905		&162.932	&\\
H$_2$CS	&$10_{0,10} -    9_{0,9}$	&342.946424		&74.132	&\\
&$10_{2,9} -    9_{2,8}$	&343.322082		&126.830	&\\
&$10_{3,8} -    9_{3,7}$	&343.409963		&192.613	&\\
&$10_{3,7} -    9_{3,6}$	&343.414146		&192.556	&\\
HCOOH   &$16_{1,16} -        15_{1,15}$	&342.521225 &127.153 \\
&$15_{1,14} -        14_{1,13}$	&343.952413 &119.774 \\ 
&$16_{0,16} -        15_{0,15}$	&345.030596 &126.494 \\
&$16_{2,15} -        15_{2,14}$	&356.137250 &141.578 

\enddata

\end{deluxetable}

\subsection{Data analysis}
\subsubsection{Methanol} \label{sec:methanol_analysis}
For the methanol data analysis, we constructed simple radiative transfer model, which uses the large velocity gradient (LVG) approximation. Dust emission and absorption within the emission region was taken into account in the way described by \citet{Sutton2004}. We assumed that the dust particles are intermixed with gas. The same physical temperature for the gas and dust components is assumed. The molecular emission region was assumed to be spherically symmetric and uniform in H$_2$ density, gas and dust temperature, gas-to-dust ratio and methanol fractional abundance. The influence of external infrared sources was not considered. The dust opacity law was chosen as $\tau_{\mathrm{dust}} \propto \lambda^{-2}$.
 We adopted a gas-to-dust mass ratio of 100 and a cross section at 1~mm of $2.6 \times 10^{-25}$~cm$^2$ \citep{Sherwood80}. In addition to the model described in \citet{Sutton2004}
 we used the collision transition rates based on the model of collisions of methanol molecules with He and para-H$_2$ molecules \citep{Cragg05}.
 Scheme of energy levels in this model includes rotational levels with the
 quantum numbers $J$ up to 22, $ \vert K\vert $ up to 9; the levels include the rotational levels of the
 ground, first and second torsionally excited states. In total, 861 levels of A-methanol and
 852 levels of E-methanol were considered according to \citet{Cragg05}.

With this model we made estimates of the hydrogen number density, $n_{\mathrm{H_2}}$,  specific column density of methanol, $N_{\mathrm{CH_3OH}} /\Delta V $, gas kinetic temperature, $T_\mathrm{kin}$ and fractional abundance of methanol $N_{\mathrm{CH_3OH}}/N_{\mathrm{H_2}}$, using the measured values of ``quasi-thermal'' (i.e. non-maser) methanol lines. The variable parameter of the source size is introduced in order to take into account beam-dilution effects. 
  
  The brightness temperatures of all detected ``quasi-thermal'' methanol lines as well as upper limits for the brightness temperatures of other lines were taken into account.

We have searched for a set of parameters which provides
 the best agreement between the values of the calculated brightness temperatures  ($T^{mod}_i$) and the measured brightness temperatures ($T^{obs}_i$). 
 This corresponds to the minimum of
  $\chi^2 = \sum_{i}^{N} ((T^{obs}_i - T^{mod}_i )/\sigma_i)^2$,
where $\sigma_i$ is observational uncertainty for a particular line.

 {The sources appear inhomogeneous in some cases. However, the LVG approximation treats the problem locally and does not actually require physical homogeneity of the source. It can be well used for inhomogeneous medium \citep[e.g.][]{Ossenkopf97}. We have chosen the simplest approximation for the case of the source
which is smaller than the beam -- introduced filling factor which is
equal to the portion of the source size in the beam. The value of the
filling factor is well constrained because the model is sensitive to
its changes. 
We do not have enough observational constraints to study real
clumpiness of the source. Introducing additional poorly constrained
parameters will reduce validity of the modelling.}

\subsubsection{Methyl cyanide} \label{sec:ch3cn_analysis}
Methyl cyanide (CH$_3$CN) is a symmetric-top molecule which is an efficient ``thermometer'' for dense molecular gas \citep[e.g.][]{Boucher80}. To derive the kinetic temperature we used the population diagram analysis which takes into account the optical depth in the CH$_3$CN lines and the beam filling factor as described by \citet{Wang10}. 

\subsection{Millimeter wave continuum}
\begin{figure*}
\begin{minipage}{0.48\textwidth}
\centering
\includegraphics[angle=-90,width=\textwidth]{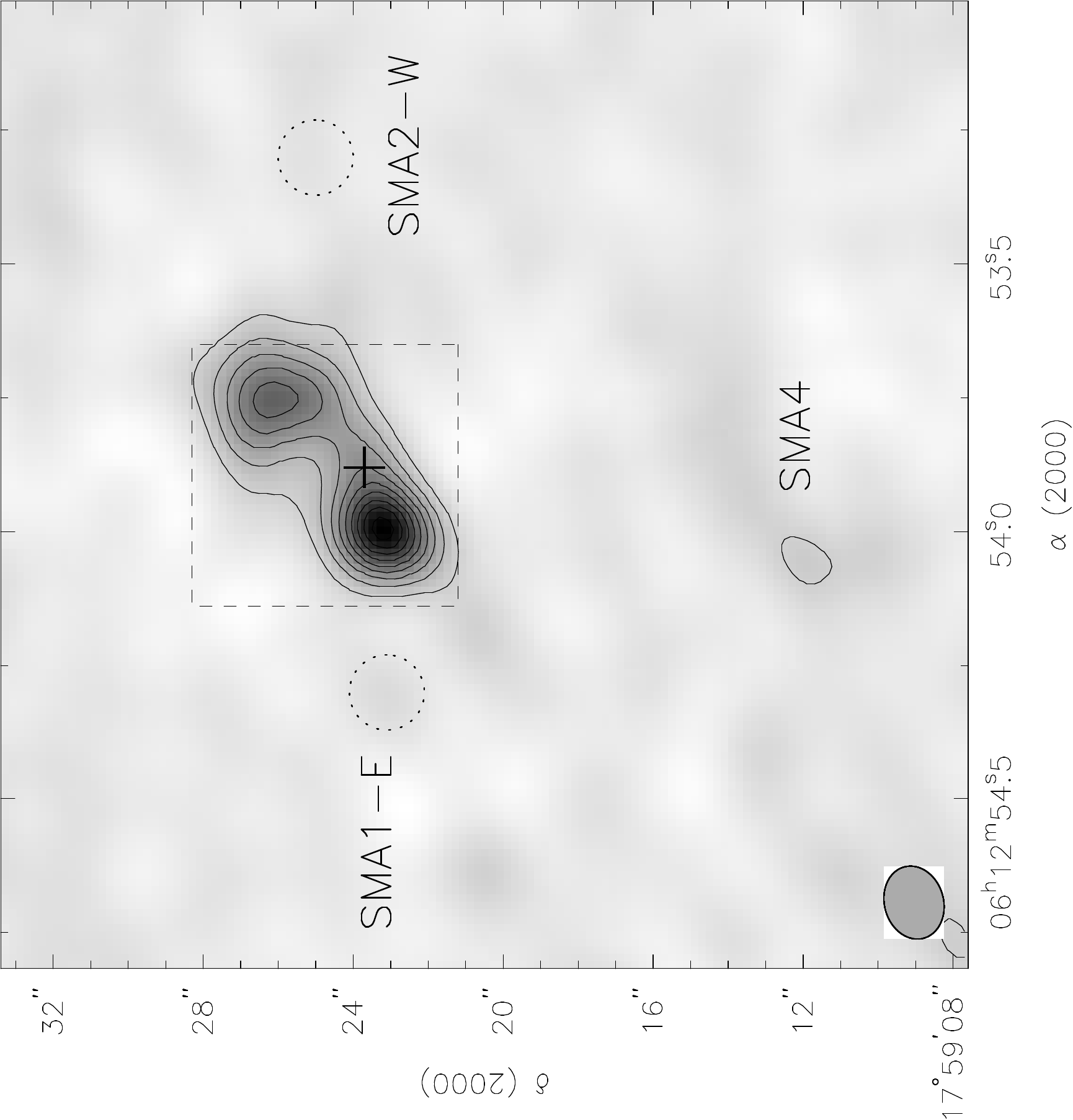}
\end{minipage}
\hfill
\begin{minipage}{0.48\textwidth}
\centering
\includegraphics[angle=-90,width=\textwidth]{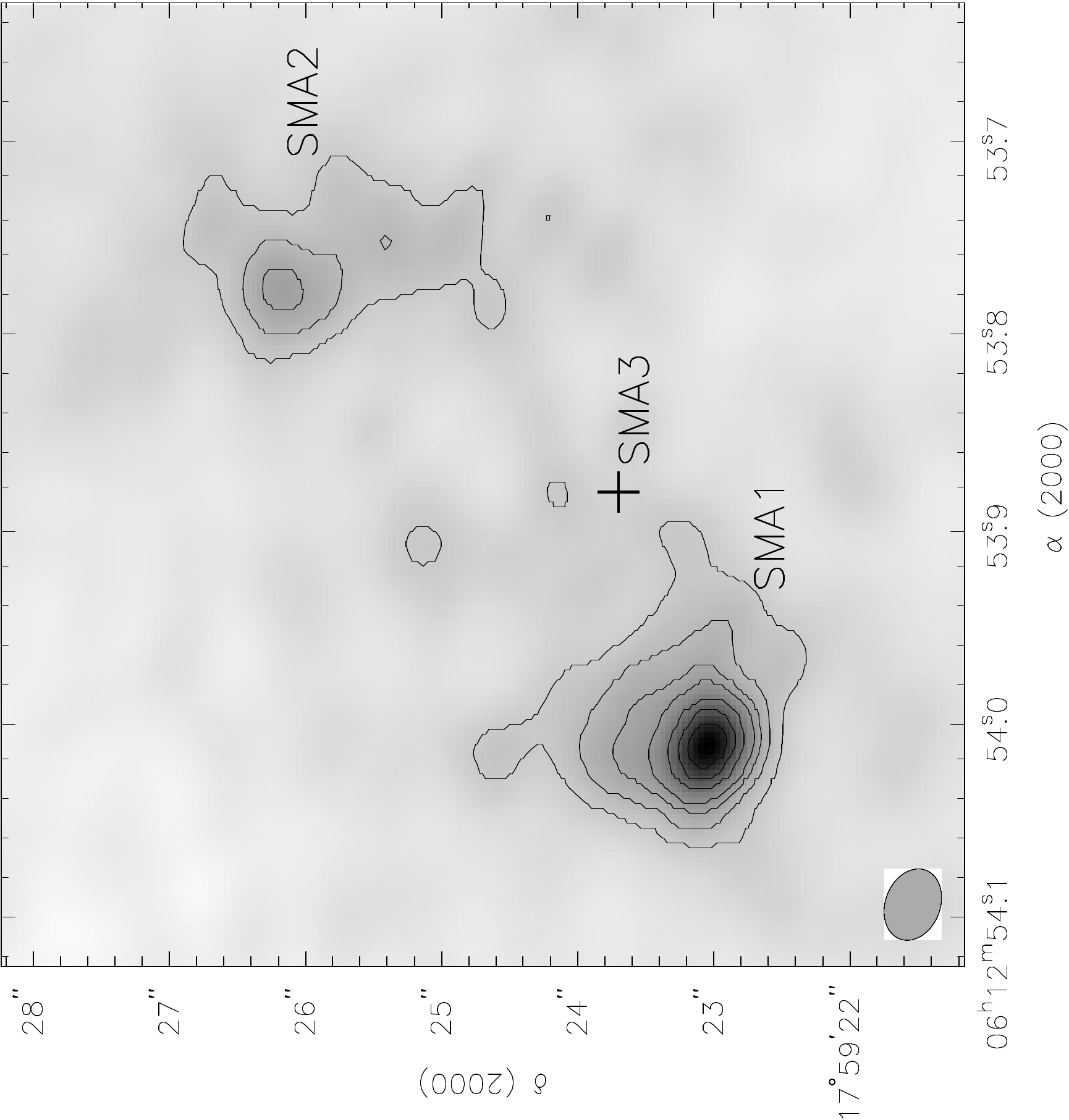}
\end{minipage}
\caption{Lef panel: map of the 0.8 mm continuum emission in the S255IR. The contour levels are (1, 2, 3, 4, 5, 6, 7, 8, 9, 10)$\times 40$~mJy\,beam$^{-1}$. The dashed rectangle indicates the area shown in the right panel.  {The dotted circles indicate positions of two molecular clumps discussed in the text.} Right panel: map of the 1.3 mm continuum emission in the S255IR. The contour levels are (2, 3, 4, 5, 7, 9, 11, 13)$\times 4$~mJy\,beam$^{-1}$. The SMA beam is shown in the lower left corner of both panels. The crosses mark the position of the SMA3 clump according to \citet{Wang11}.}
\label{fig:cont}
\end{figure*}

Maps of the continuum emission at 0.8 and 1.3~mm are presented in Fig.~\ref{fig:cont}. The 1.3~mm map was obtained by combining our new measurements in the very extended configuration with the previous measurements in the compact configuration (Paper~I). It shows a rather extended emission and compact cores in the SMA1 and SMA2 clumps.

Estimates of the continuum source parameters are given in Tables~\ref{table:cont-350},\ref{table:cont-230}. For the measurements at 1.3~mm we indicate the parameters of the compact cores seen in the very  {extended} configuration. Parameters of the more extended components were presented in Paper~I. The measured fluxes are several hundreds mJy at 350~GHz and several tens mJy at 225~GHz in the very extended configuration. The deconvolved sizes are from $ \sim 1 $ arcsecond to a few arcseconds at 345~GHz and about 0.3 arcsecond for the compact cores at 225~GHz.

\begin{deluxetable*}{lllcccr}
\tablecaption{Names, positions, flux densities, deconvolved angular sizes and position angles of the millimeter wave continuum sources measured at 350~GHz in the compact configuration. \label{table:cont-350}}
\tablehead{\colhead{Name} &\colhead{$\alpha$(2000)} &\colhead{$\delta$(2000)} &\colhead{$S_{350}$} &\colhead{$\theta_{\mathrm{max}}$} &\colhead{$\theta_{\mathrm{min}}$} &\colhead{P.A.} \\ \colhead{ } &\colhead{(h m s)} &\colhead{($^\circ$ $^\prime$ $^{\prime\prime}$)} &\colhead{(Jy)} &\colhead{($^{\prime\prime}$)} &\colhead{($^{\prime\prime}$)} &\colhead{($^\circ$)}}
\tablecolumns{8}
\startdata
S255IR-SMA1 &6:12:54.00 &17:59:23.2 &0.50  &1.4    &0.6  &--11  \\
S255IR-SMA2 &6:12:53.76 &17:59:26.1 &0.56  &2.4    &2.1  &4  \\
S255IR-SMA3 &6:12:53.86 &17:59:23.7 &0.13  &1.3    &0.3  &44 \\
S255IR-SMA4 &6:12:54.01 &17:59:12.0 &0.25  &5.9    &3.8  &--82 
\enddata

\end{deluxetable*}

\begin{deluxetable*}{lllcccr}
\tablecaption{Names, positions, flux densities, deconvolved angular sizes and position angles of the millimeter wave continuum sources measured at 225~GHz in the very extended configuration. \label{table:cont-230}}
\tablehead{\colhead{Name} &\colhead{$\alpha$(2000)} &\colhead{$\delta$(2000)} &\colhead{$S_{225}$} &\colhead{$\theta_{\mathrm{max}}$} &\colhead{$\theta_{\mathrm{min}}$} &\colhead{P.A.} \\ \colhead{ } &\colhead{(h m s)} &\colhead{($^\circ$ $^\prime$ $^{\prime\prime}$)} &\colhead{(Jy)} &\colhead{($^{\prime\prime}$)} &\colhead{($^{\prime\prime}$)} &\colhead{($^\circ$)}}
\tablecolumns{8}
\startdata
S255IR-SMA1 &6:12:54.010 &17:59:23.06 &0.058  &0.30    &0.27  &--19  \\
S255IR-SMA2 &6:12:53.779 &17:59:26.16 &0.016  &0.49    &0.31  &--27 
\enddata

\end{deluxetable*}

\subsection{Basic properties of the molecular emission}

Here we give the general description of the observed molecular emission. A more detailed information on the relevant species and transitions is presented in the following sections.

\begin{figure*}
\centering
\includegraphics[angle=-90,width=\textwidth]{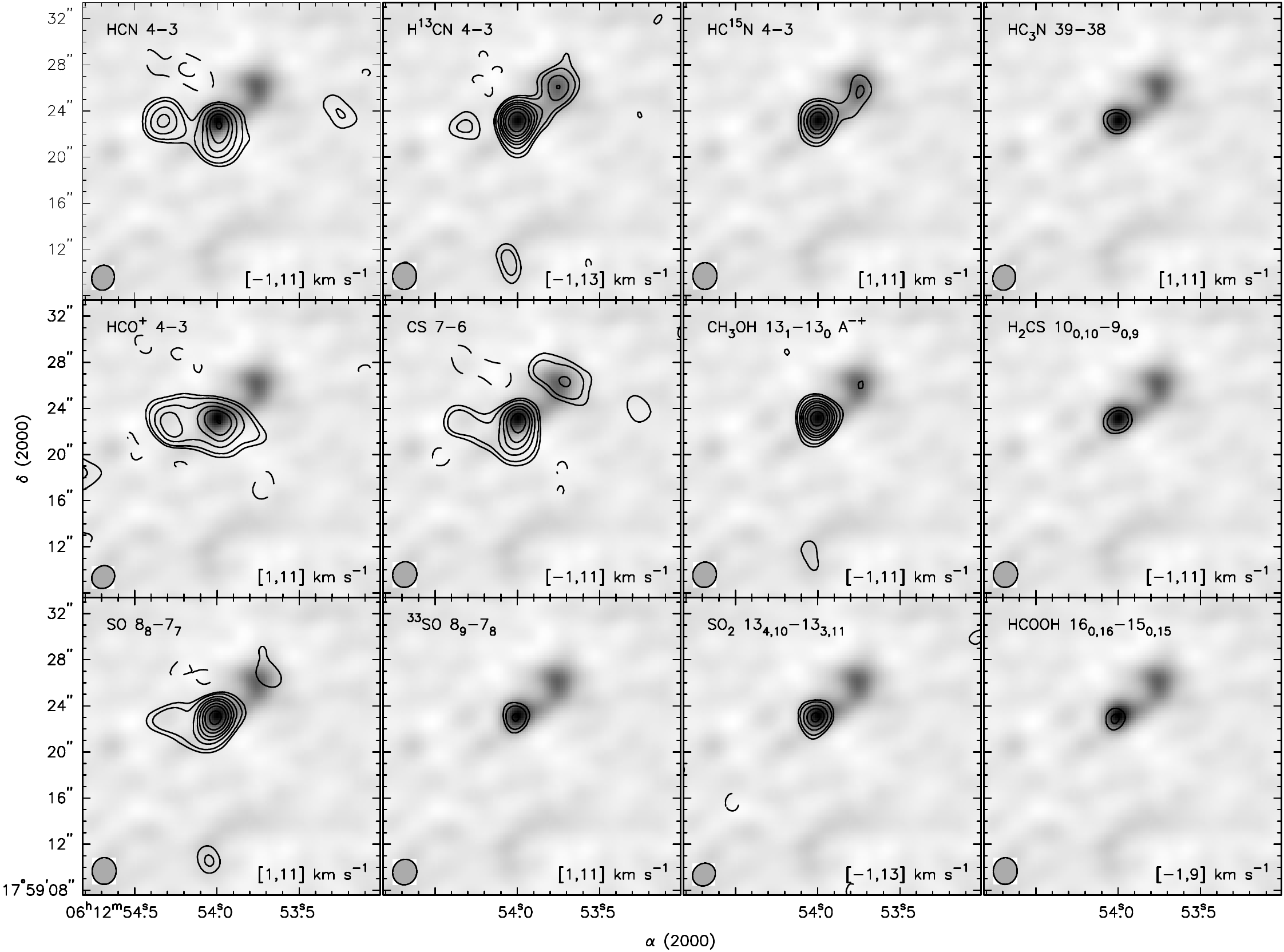}
\caption{Maps of the integrated line emission for several representative transitions of various molecules (contours) overlaid on the image of the 0.8~mm continuum emission. The velocity interval is shown at the bottom of each panel. The dashed contours show negative features due to the missing flux. The contour levels are the following: (--3, --2, 2, 3, 5, 7, 9, 12, 15, 20)$\times A$~Jy\,beam$^{-1}$\,km\,s$^{-1}$, where $ A=3 $ for HCN and HCO$^+$, $ A=1 $ for H$^{13}$CN, HC$^{15}$N, HC$_3$N, H$_2$CS, CH$_3$OH, $^{33}$SO, SO$_2$, and HCOOH, $ A=2 $ for CS, and $ A=1.5 $ for SO.  {The SMA beam is shown in the lower left corner of each panel.}}
\label{fig:4x3maps}
\end{figure*}

The general morphology and kinematics of the molecular emission was described in Paper~I. Our new data set contains several tens of molecular transitions. In Fig.~\ref{fig:4x3maps} we present representative maps obtained with the SMA at 0.8~mm.

These maps confirm that SMA1 is the brightest source of molecular emission in this area. High-excitation lines are observed exclusively towards this clump. Several lower excitation molecular lines are detected also in SMA2. Emission of HCN, SO and CH$_3$OH is observed also in SMA4. HCN and CS emission is probably present in the area to the west of SMA2, designated as S255IR-N$_2$H$^+$(1) in Paper~I.  {Since it is observed not in N$_2$H$^+$ only and for simplicity we shall designate it hereafter as SMA2-W}.

A new feature, not noticed in our Paper~I and in other previous works is a molecular clump to the east of SMA1 with a rather strong emission in the HCN, HCO$^+$, CS and SO lines (Fig.~\ref{fig:4x3maps}). We shall designate it as SMA1-E. This clump is located near the head of the jet observed in the NIR emission (see below). 

\begin{figure*}
\centering
\includegraphics[width=\textwidth]{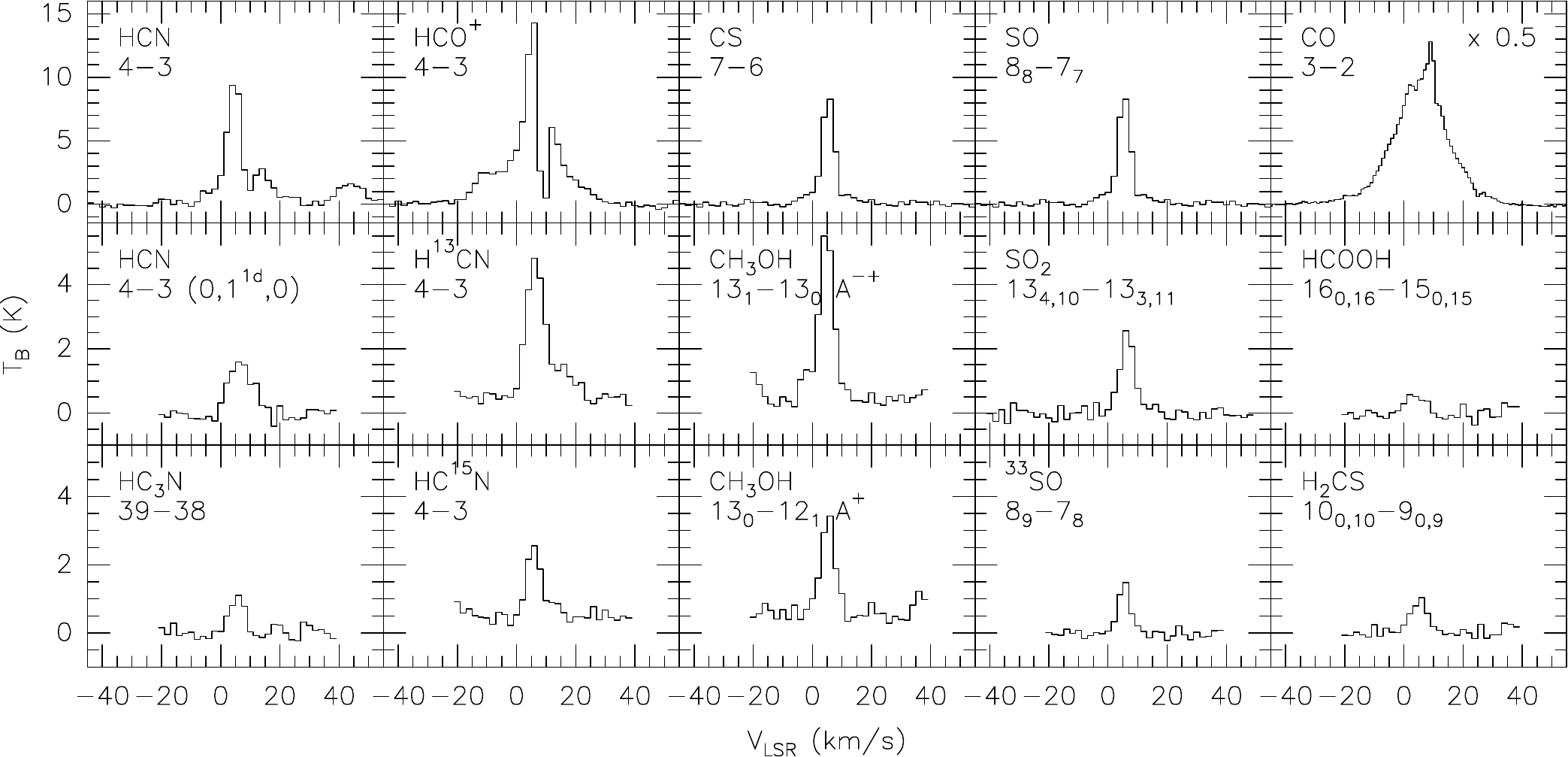}
\caption{Spectra of several representative molecular transitions towards the SMA1 clump. The CO(3--2) spectrum is scaled by the factor of 0.5.  {It is obtained from the combined SMA and IRAM 30m data. The other spectra are from the SMA data only. The angular resolution is about 2 arcseconds.}}
\label{fig:sma1-spectra}
\end{figure*}

\begin{figure*}
\centering
\includegraphics[width=\textwidth]{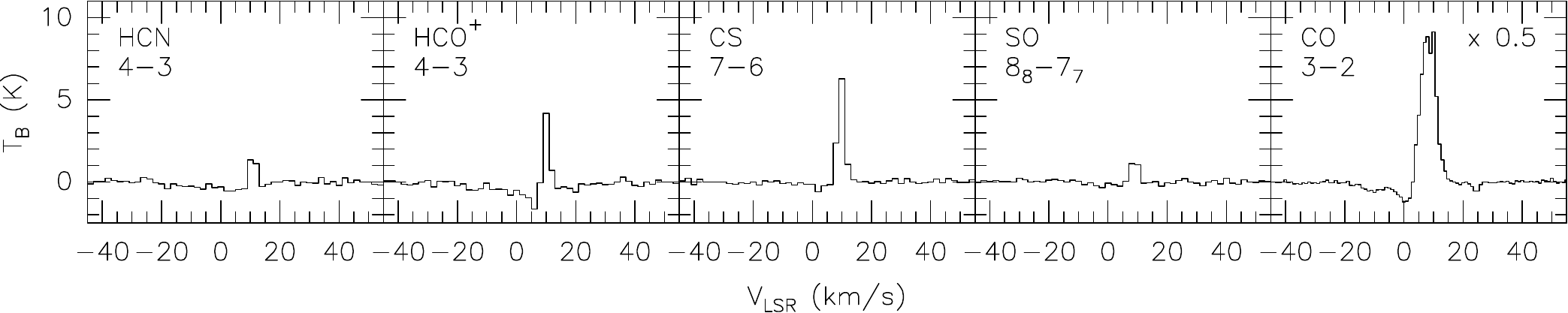}
\caption{Spectra of several representative molecular transitions towards the SMA2 clump. The CO(3--2) spectrum is scaled by the factor of 0.5.  {It is obtained from the combined SMA and IRAM 30m data. The other spectra are from the SMA data only. The angular resolution is about 2 arcseconds.}}
\label{fig:sma2-spectra}
\end{figure*}

Spectra of several representative transitions towards the SMA1 and SMA2 clumps are presented in Figs.~\ref{fig:sma1-spectra},\ref{fig:sma2-spectra}.

As mentioned in Paper~I the main emission peak toward the SMA1 clump is at $ V_\mathrm{LSR} \sim 4-5 $~km\,s$^{-1}$. The line width is $ \ga 5 $~km\,s$^{-1}$. Several lines including CO, HCN, HCO$^+$, CS, SO show broad wings indicative of high velocity outflow. We discuss this feature in the following sections. A rather strong emission is detected from vibrationally excited HCN with the excitation energy $ \sim 1000 $~K above the ground level. 

The line emission from the SMA1-E clump peaks at about 8~km\,s$^{-1}$. The linewidth is large, $ \sim 5 $~km\,s$^{-1}$.

The emission from the SMA2 clump is observed at about 10~km\,s$^{-1}$. The lines are narrow, $ \sim 2 $~km\,s$^{-1}$.

With the very extended array we detected the line emission almost exclusively from the SMA1 compact core. No emission was detected in C$^{18}$O and SiO indicating an absence of compact structures in the lines of these molecules. The results of these observations are presented and discussed below.

\section{Structure, kinematics and physical properties of dense cores}
Our data presented in Paper~I and here as well as data by \citet{Wang11} indicate the presence of four continuum clumps in the S255IR area, designated from S255IR-SMA1 to S255IR-SMA4. In addition, in Paper~I we detected clumps with a rather strong molecular emission without continuum counterpart in the SMA data. One of them, S255IR-N$_2$H$^+$(1), is located close to SMA2 and was observed in the N$_2$H$^+$, NH$_3$ and several CH$_3$OH lines. Estimates of their basic physical properties were presented in Paper~I. Here we investigate further these objects using the new data set.

\subsection{SMA1}
SMA1 is the brightest object in this area. As shown in Sect.~\ref{sec:results} most of the observed molecular lines are detected only here. The deconvolved size of the continuum source at 350~GHz measured with the SMA in the compact configuration (Table~\ref{table:cont-350}) is close to that found in Paper~I. The flux density measured at 350~GHz is only slightly higher than the flux density obtained at 284~GHz (Paper~I). It is worth noting that in Paper~I we could not separate the SMA1 and SMA3 clumps. However, even if we take the integrated flux of these clumps at 350~GHz, the spectral index in the range $ 284-350 $~GHz will be only 1.6 which is inconsistent with a presumably optically thin  {dust emission at these frequencies}. Most probably this implies a significantly larger flux loss at 350~GHz in comparison with the measurements at 284~GHz due to a smaller beam size. 

The deconvolved size of the continuum source detected with the SMA in the very extended configuration is about 0$\farcs$3 (Table~\ref{table:cont-230}) which corresponds to about 500~AU. The observed aspect ratio for this core is close to unity. 

\subsubsection{Kinematics}
\begin{figure*}
\begin{minipage}{0.48\textwidth}
\centering
\includegraphics[width=\columnwidth]{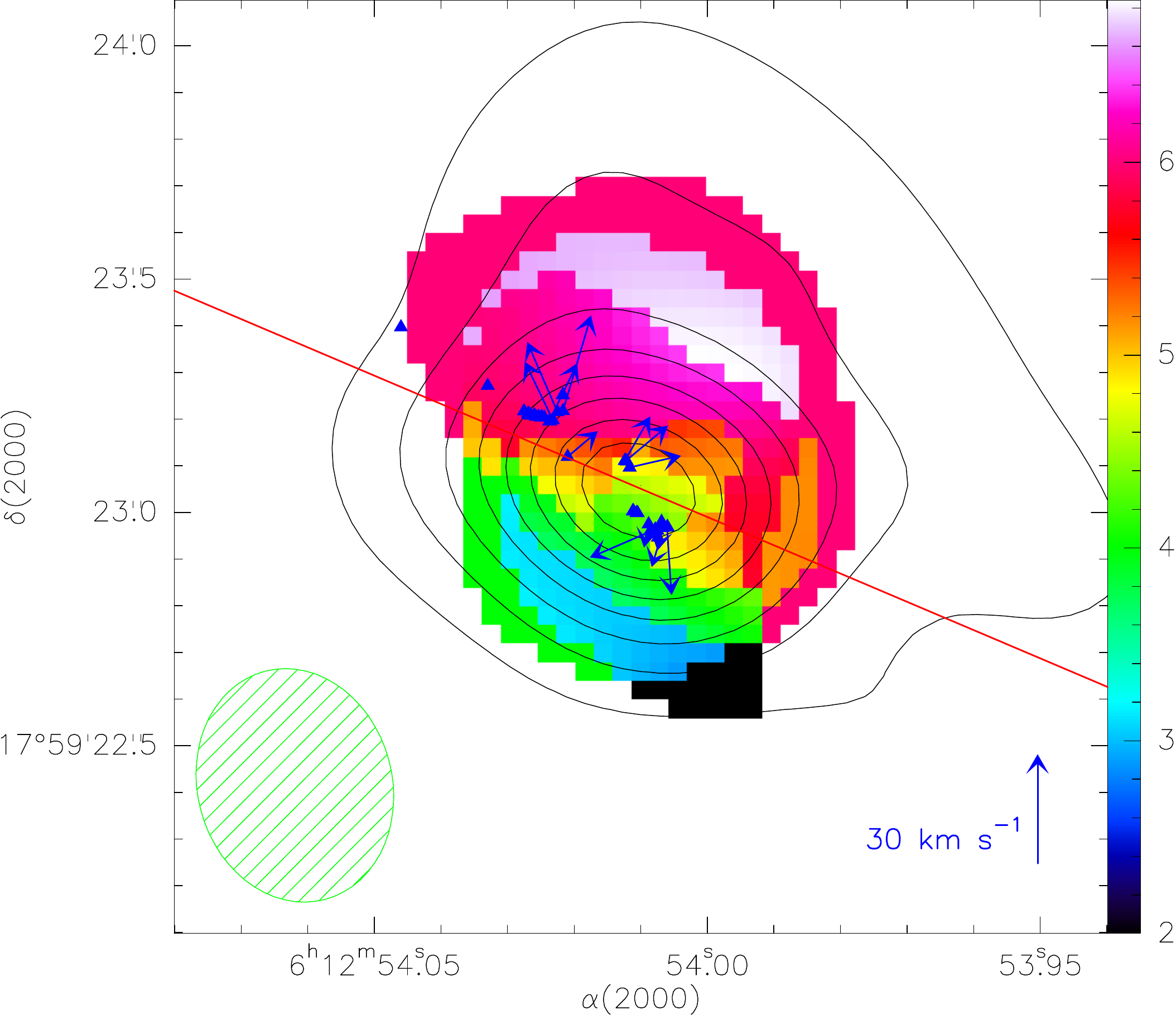}
\end{minipage}
\hfill
\begin{minipage}{0.48\textwidth}
\centering
\includegraphics[width=\columnwidth]{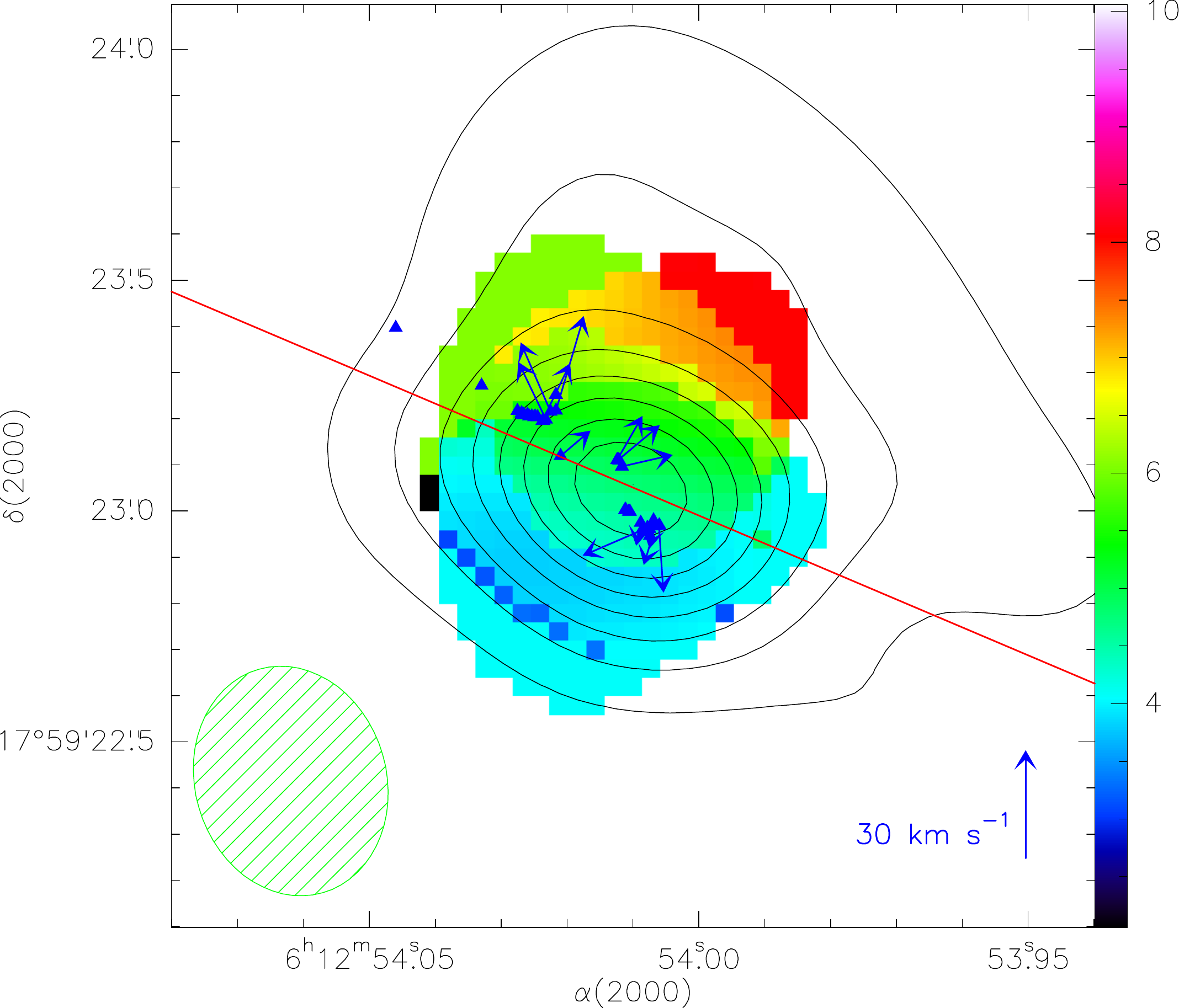}
\end{minipage}
\caption{Maps of the first moment of the CH$_3$OH emission in the $4_{2} - 3_{1}$	E line (left panel, color scale) and CH$_3$CN emission in the $12_{3} - 11_{3}$ line (right panel, color scale) towards the SMA1 clump. The intensity cut-off is 100~mJy\,beam$^{-1}$\,km\,s$^{-1}$ for CH$_3$OH and 200~mJy\,beam$^{-1}$\,km\,s$^{-1}$ for CH$_3$CN. Contours show the 1.3~mm continuum emission. The triangles mark water masers \citep{Goddi07}  {and the arrows indicate their proper motions (in the cases when they are measured)}. The red line indicates the jet axis as found by \citet{Howard97}.  {The beam for the molecular maps is shown in the lower left corner of both panels. The scale for the velocities of the proper motions is shown in the lower right corner. The pixels with the lowest velocities appear as the blue squares in the right panel.}}
\label{fig:sma1_ch3oh-vel}
\end{figure*}

\citet{Wang11} noticed rotation of the core around the axis roughly parallel to the outflow direction. Now, at sub-arcsecond resolution we can investigate the core kinematics on smaller scales. In Fig.~\ref{fig:sma1_ch3oh-vel} we present maps of the first moment of the CH$_3$OH emission in the $4_{2} - 3_{1}$	E line and CH$_3$CN emission in the $12_{3} - 11_{3}$ line. In this figure we also indicate the axis of the jet  {previously identified through} IR observations \citep{Howard97} and locations of the water masers measured with the VLBA \citep{Goddi07}.

\begin{figure}
\centering
\includegraphics[angle=-90, width=\columnwidth]{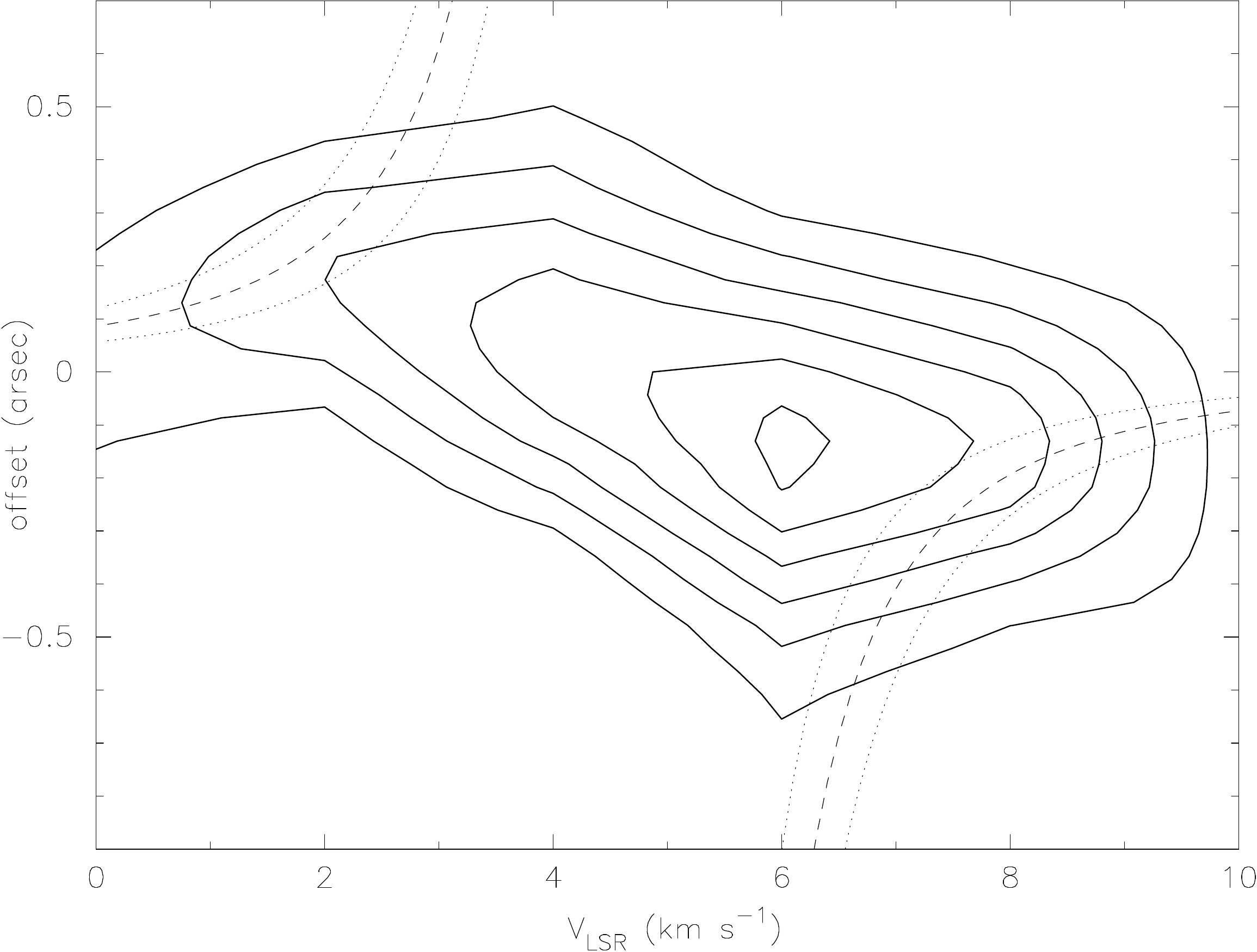}
\caption{The position-velocity diagram for the CH$_3$OH emission in the $4_{2} - 3_{1}$	E line along the cut through the SMA1 center perpendicular to the jet axis  {(PA = 157$^\circ$)}. The dashed curves correspond to Keplerian rotation around a central mass of 20~M$_\odot$ with the inclination angle of 25$^\circ$, the dotted curves correspond to the inclination angles of 25$^\circ \pm 5^\circ$.}
\label{fig:sma1_ch3oh-pv}
\end{figure}

This figure clearly shows that the core is really rotating around the axis of the jet. The rotation velocities along the line of sight  {amount to} a few km\,s$^{-1}$  {(it varies from about 3~km\,s$^{-1}$ to about 7~km\,s$^{-1}$ for CH$_3$OH $4_{2} - 3_{1}$~E and to about 8~km\,s$^{-1}$ for CH$_3$CN $12_{3} - 11_{3}$)}. In Fig.~\ref{fig:sma1_ch3oh-pv} we present the position-velocity diagram for the CH$_3$OH emission in the $4_{2} - 3_{1}$ E line along the cut through the core center perpendicular to the jet axis  {(PA = 157$^\circ$)}. It clearly shows the velocity gradient  {which can be consistent with} Keplerian rotation but the angular resolution is still insufficient for firm conclusions about the character of this rotation.

\begin{figure}
\centering
\includegraphics[width=\columnwidth]{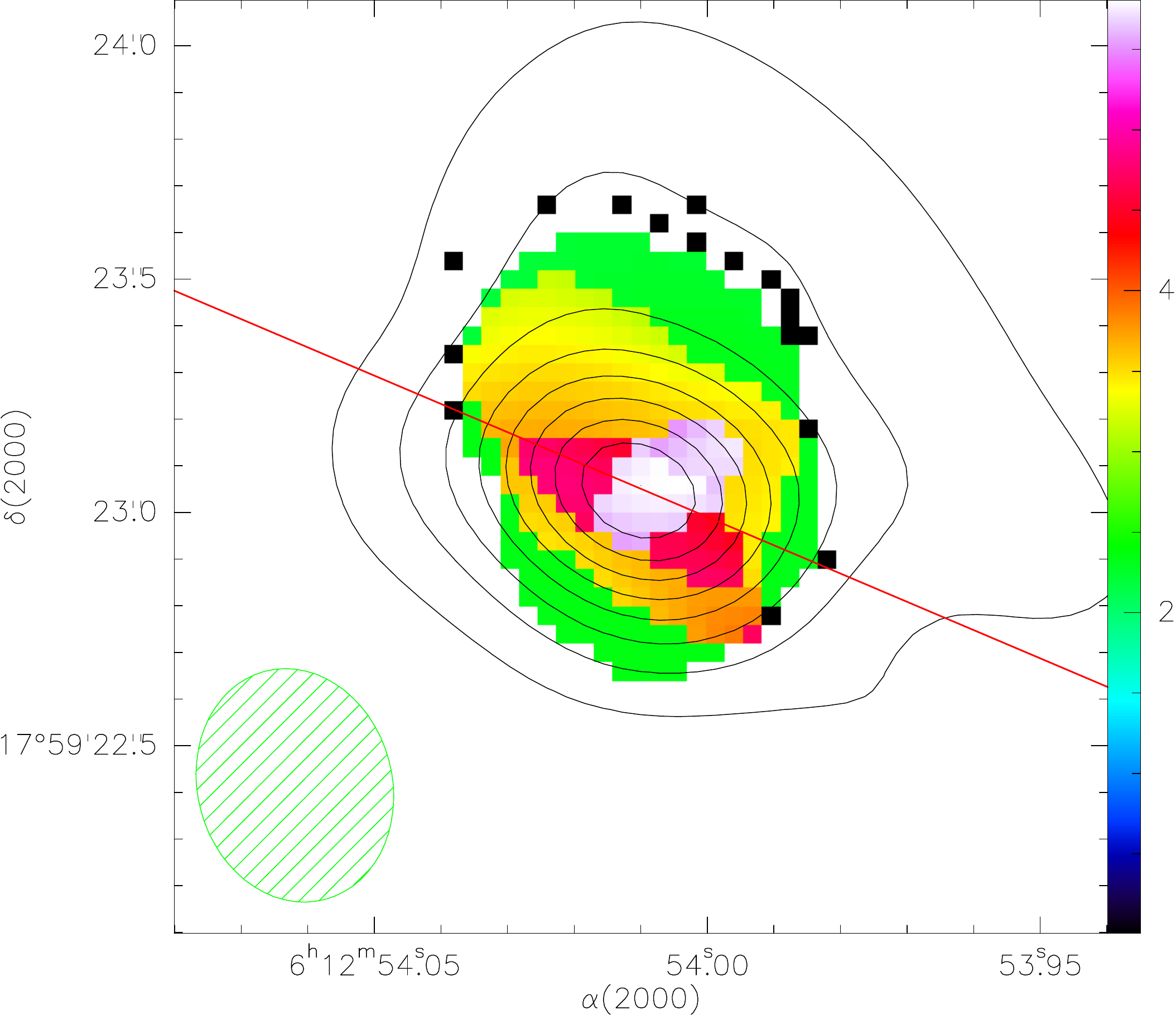}
\caption{Map of the second moment of the CH$_3$OH emission in the $4_{2} - 3_{1}$	E line (color scale). Contours show the 1.3~mm continuum emission.  {The beam for the molecular map is shown in the lower left corner.}}
\label{fig:sma1_ch3oh-width}
\end{figure}

An implicit indication of a probable further increase of the rotation velocity in the innermost part of the core comes from the line width map (Fig.~\ref{fig:sma1_ch3oh-width}). It shows a significant increase of the line width in the center which can be related to rotation  {(the line width increases from about 2~km\,s$^{-1}$ at the periphery to about 6~km\,s$^{-1}$ in the center)}.

It is also worth mentioning an increased line width along the jet axis (Fig.~\ref{fig:sma1_ch3oh-width}). It can be probably explained by an increased turbulence caused by the passage of the jet. 

\begin{figure}
\centering
\includegraphics[width=\columnwidth]{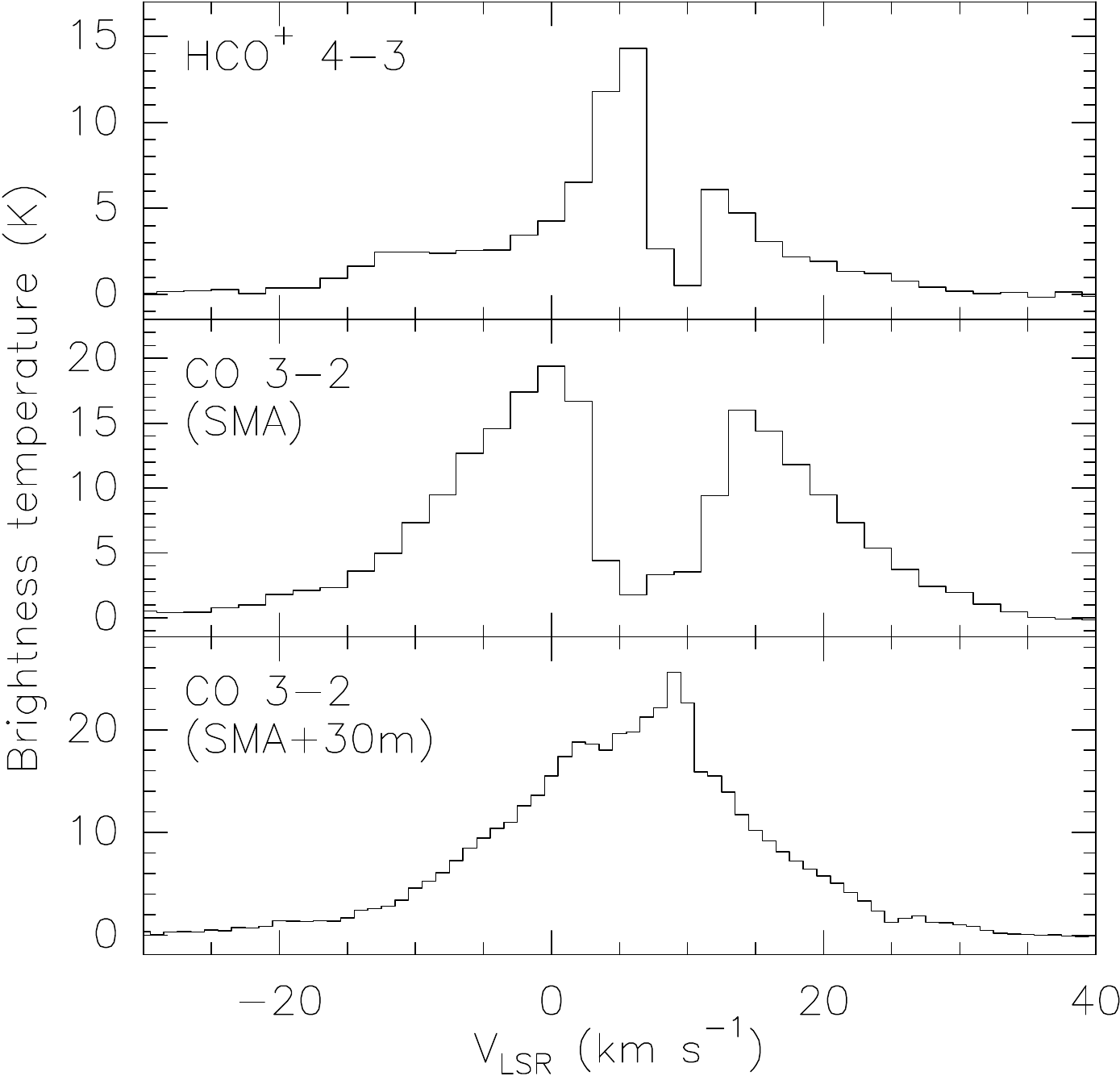}
\caption{ {The SMA spectra of HCO$^+$(4--3) and CO(3--2) as well as the combined SMA+30m spectrum of CO(3--2) towards SMA1. The angular resolution is about 2 arcseconds.}}
\label{fig:sma1_abs}
\end{figure}

 {The HCO$^+$(4--3) and probably HCN(4--3) spectra towards SMA1 (Fig.~\ref{fig:sma1-spectra}) show the red-shifted absorption dip which may be suggestive of infall. The CO(3--2) spectrum measured with the SMA has a similar (although broader) feature (Fig.~\ref{fig:sma1_abs}). At the same time the CO(3--2) spectrum obtained from the combined SMA and IRAM-30m data shows a peak at these velocities. Most probably it means that this feature is related to an extended component resolved out by the SMA, although an infall cannot be fully excluded.}

\subsubsection{Physical properties}
\begin{figure}
\centering
\includegraphics[width=\columnwidth]{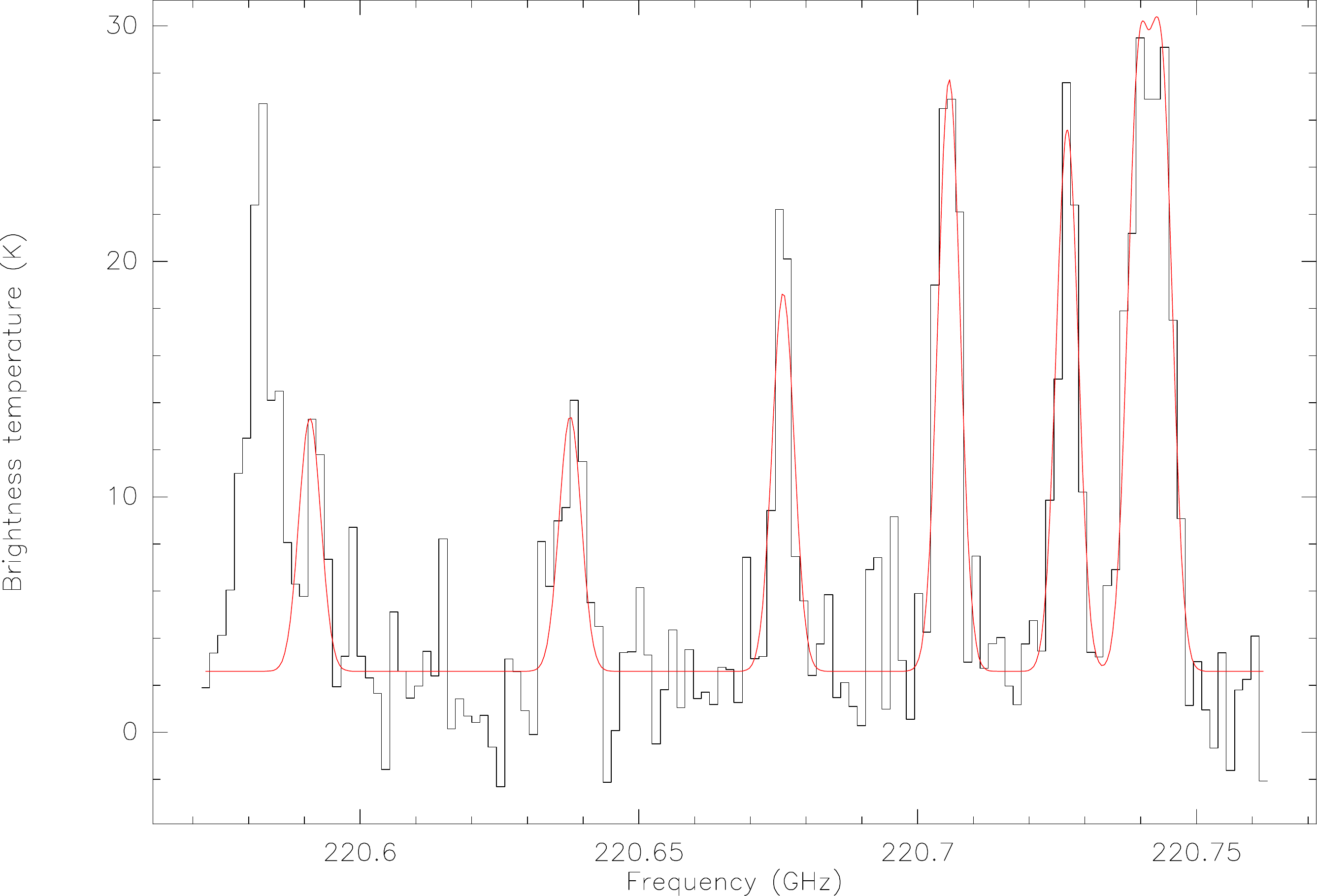}
\caption{The CH$_3$CN spectrum measured towards the SMA1 emission peak and the model fit.}
\label{fig:sma1_ch3cn-fit}
\end{figure}

Let us consider physical parameters of this core.
We estimated gas kinetic temperature from the CH$_3$CN and CH$_3$OH observations. Modeling of the CH$_3$CN emission (Fig.~\ref{fig:sma1_ch3cn-fit}) as described in Sect.~\ref{sec:ch3cn_analysis} yields kinetic temperature in the range (74.9 -- 197.6)~K (1$\sigma$ confidence interval) towards the emission peak (Table~\ref{table:ch3cn-sma1}).  {We have} detected a large number (15) of CH$_3$OH transitions towards SMA1 (Table~\ref{table:lines-vext}). From the CH$_3$OH analysis (Sect.~\ref{sec:methanol_analysis}) we obtain $ T_\mathrm{kin} = 178 $~K with the $165-195$~K 1$\sigma$ confidence interval at the center, $ T_\mathrm{kin} = 170-200 $~K at 0$\farcs$2 to the north and $ T_\mathrm{kin} = 140-165 $~K at 0$\farcs$2 to the south (Table~\ref{table:ch3oh-sma1}).  {Therefore there is probably a temperature gradient in the north-south direction. The temperature seems to be anti-correlated with the CH$_3$OH column density (Table~\ref{table:ch3oh-sma1}).} 

\begin{deluxetable}{lc}
\tablecaption{ {Kinetic temperature, CH$_3$CN column density, beam filling factor and line optical depth derived from the CH$_3$CN data obtained in the very extended configuration (HPBW $ \approx 0\farcs4 $) at the central position in SMA1. The 1$\sigma$ confidence intervals are indicated in the parentheses.}\label{table:ch3cn-sma1}}
\tablehead{\colhead{Parameter} &\colhead{Fit results} 
}
\tablecolumns{2}
\startdata
$T_\mathrm{kin}$ (K) &128.8         (74.9 -- 197.6)\\
$N_\mathrm{L}(\mathrm{CH_3CN})$ ($ 10^{16} $~cm$^{-2}$)  & 2.12   (0.96 -- 5.80)\\
Beam filling factor  &0.23            (0.18 -- 0.36)\\
$\tau(K=0)$ 	&2.53		(0.93 -- 8.37)
\enddata
\end{deluxetable}

\begin{figure}
\centering
\includegraphics[width=\columnwidth]{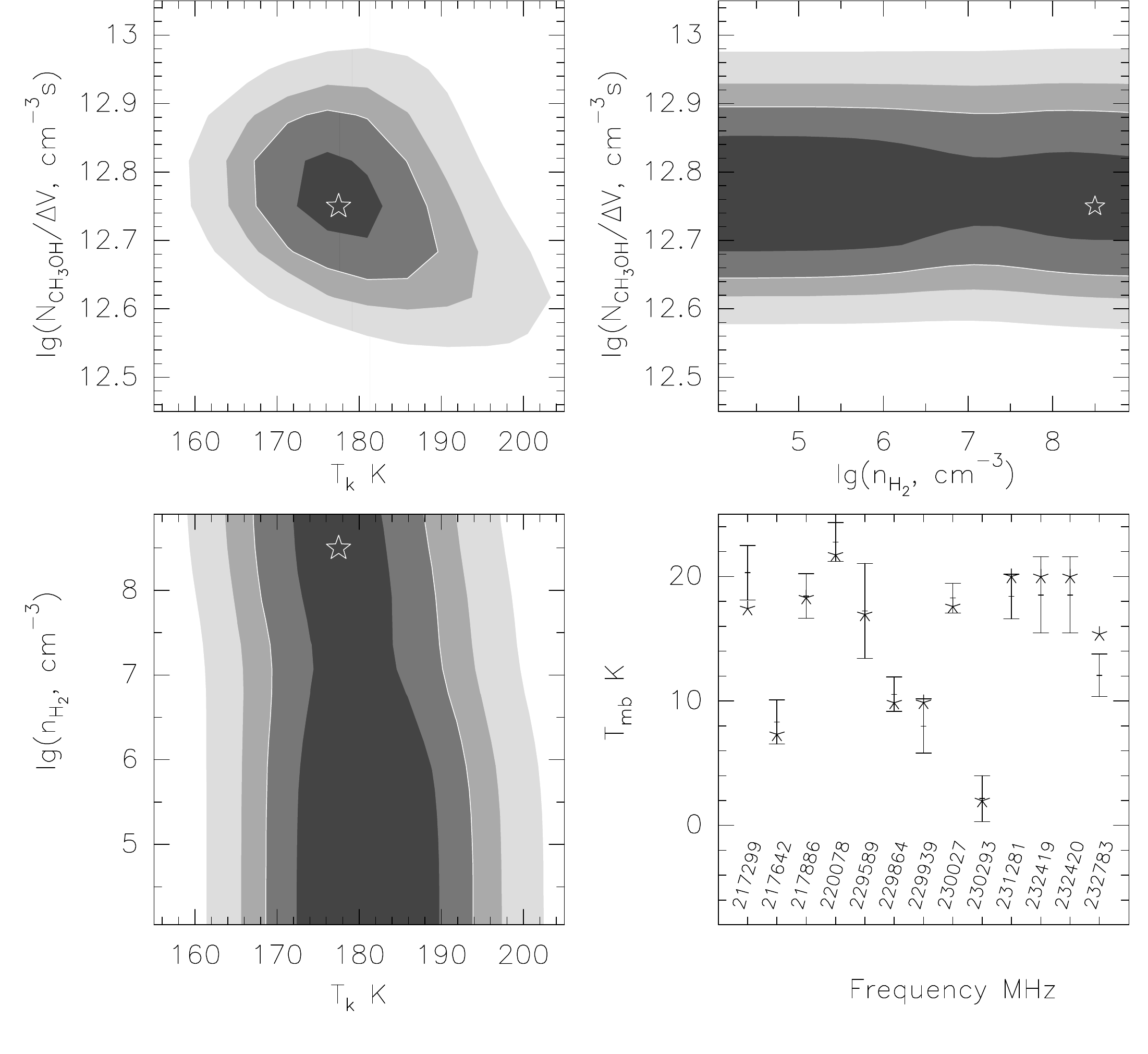}
\caption{ {An example of the CH$_3$OH data fitting for the central position in SMA1. Variations of $\chi^2$ with the model parameters are shown in the  2 top and left bottom panels. 
The minimum of $\chi^2$ is marked by the star. Confidence intervals of 0.25, 1, 1.6 2.6 $\sigma$ are plotted on the grey scale.  
The confidence interval of 1 $\sigma$ is marked by the white line.  
The comparison of the model (stars) and observed main-beam temperatures is shown in the right bottom panel.  
}}
\label{fig:sma1_ch3oh-fit}
\end{figure}


\begin{deluxetable*}{cccccc}
\tablecaption{Physical parameters derived from the CH$_3$OH data obtained in the very extended configuration (HPBW $ \approx 0\farcs4 $) at several positions in SMA1. The 1$\sigma$ confidence intervals are indicated in the parentheses (the density $n_{\rm H_2}$ is not constrained by the model as evident from Fig.~\ref{fig:sma1_ch3oh-fit}). \label{table:ch3oh-sma1}}
\tablehead{\colhead{Offsets} &\colhead{$T_\mathrm{kin}$} &\colhead{$N_{\rm{CH_3OH}}/{\rm\Delta}V  $} &\colhead{$n_{\rm H_2}$} &\colhead{beam filling factor} &\colhead{$N_{\rm{CH_3OH}}/N_{\rm H_2}$}\\ 
\colhead{($^{\prime\prime}$, $^{\prime\prime}$)} &\colhead{(K)} &\colhead{($10^{12}$ cm$^{-3}$s)} &\colhead{($10^8$ cm$^{-3}$)} &\colhead{(\%)} &\colhead{ }
}
\tablecolumns{6}
\startdata
 0, +0.2 & 183 (170---200)&	3.6 (2.8---5.0)  &  3.2   &  14.8  (13.9---16.4) &  $10^{-6} (>10^{-7})$ \\
 0, 0             & 178 (165---195)&	5.6 (4.0---9.5)  &  3.2   &  16.0 (15.2---17.2)  &  $10^{-6} (>10^{-7})$ \\
0, $-$0.2 & 153 (140---165)&	8.9 (5.0---12.6) &  0.2   &  15.2  (13.3---16.0)  &  $10^{-6} (>10^{-7})$
\enddata

\end{deluxetable*}

 {We adopt the gas kinetic temperature of 170~K as the average value (it is close to the weighted average of the CH$_3$OH and CH$_3$CN results).} Assuming the same temperature (170~K) for the dust we obtain a total mass of this hot component of about 0.3~M$_\odot$. As in Paper~I we assume a gas-to-dust mass ratio of 100, and adopt a dust absorption efficiency following \citet{Ossenkopf94}. The peak gas column density estimated from the continuum data is $ N(\mathrm{H_2}) \sim 3\times 10^{24} $~cm$^{-2}$.  {With a} size of 500~AU the mean density of hot gas is about $6\times 10^8$~cm$^{-3}$. This is 2 times higher than our estimate of the SMA1 mean density in Paper~I taking into account the difference in the adopted distances here and in Paper~I. This increase  {can be expected} since now we consider much smaller scales near the core center. The CH$_3$OH modeling puts no significant constraints on density. The relative abundance of CH$_3$OH is $ \sim 10^{-6} $.

Both CH$_3$CN and CH$_3$OH observations indicate beam filling factor of 0.15--0.2. This means that the source is very inhomogeneous on the 0$\farcs$4 scale and probably consists of clumps with the size $ \ll 500 $~AU and density much higher than the mean density found above. This higher density estimate does not contradict the CH$_3$OH data. A similar picture of clumpy medium was inferred from our observations of high mass star forming regions on larger scales \citep{Pirogov08,Pirogov12}. Most probably it reflects turbulence in the cores.

\begin{figure}
\centering
\includegraphics[width=\columnwidth]{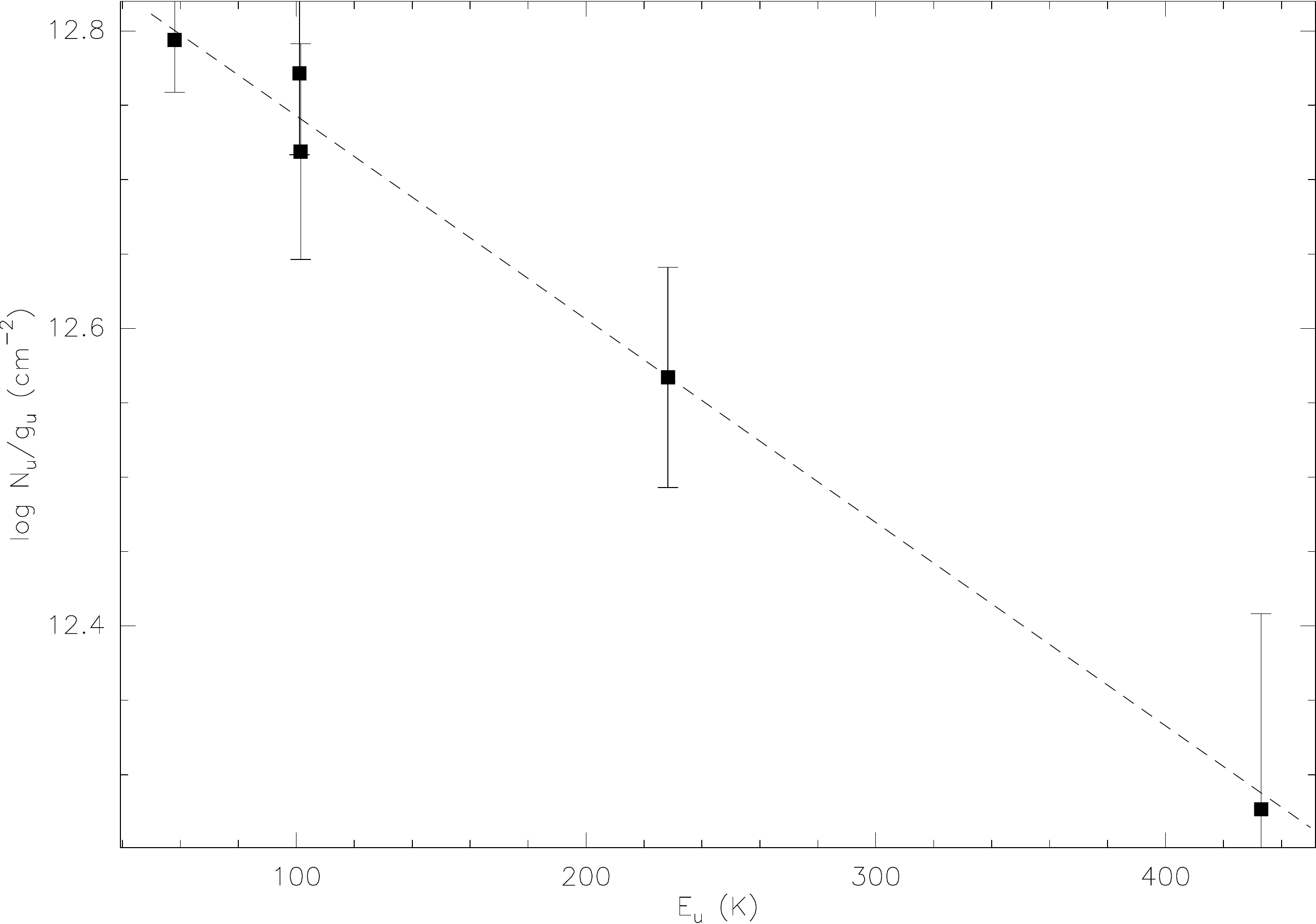}
\caption{The rotational diagram for HNCO towards the emission peak.}
\label{fig:hnco-rd}
\end{figure}

Our data set includes several other tracers of the hot gas. One of the most important is HNCO. We detected HNCO lines in different $K_{-1}$ ladders ($ K_{-1} = 0, 1, 2, 3 $) with the excitation energies up to $ \sim 400 $~K (Table~\ref{table:lines-vext}). In Fig.~\ref{fig:hnco-rd} we present the rotational diagram for the observed HNCO transitions. This diagram was obtained from peak integrated line intensities found by 2D Gaussian fitting of the integrated intensity maps in different HNCO lines. We used the line strengths and dipole moment from the Cologne Database for Molecular Spectroscopy \citep{Mueller01,Mueller05}. 

\begin{figure*}
\begin{minipage}{0.49\textwidth}
\centering
    \includegraphics[width=\columnwidth]{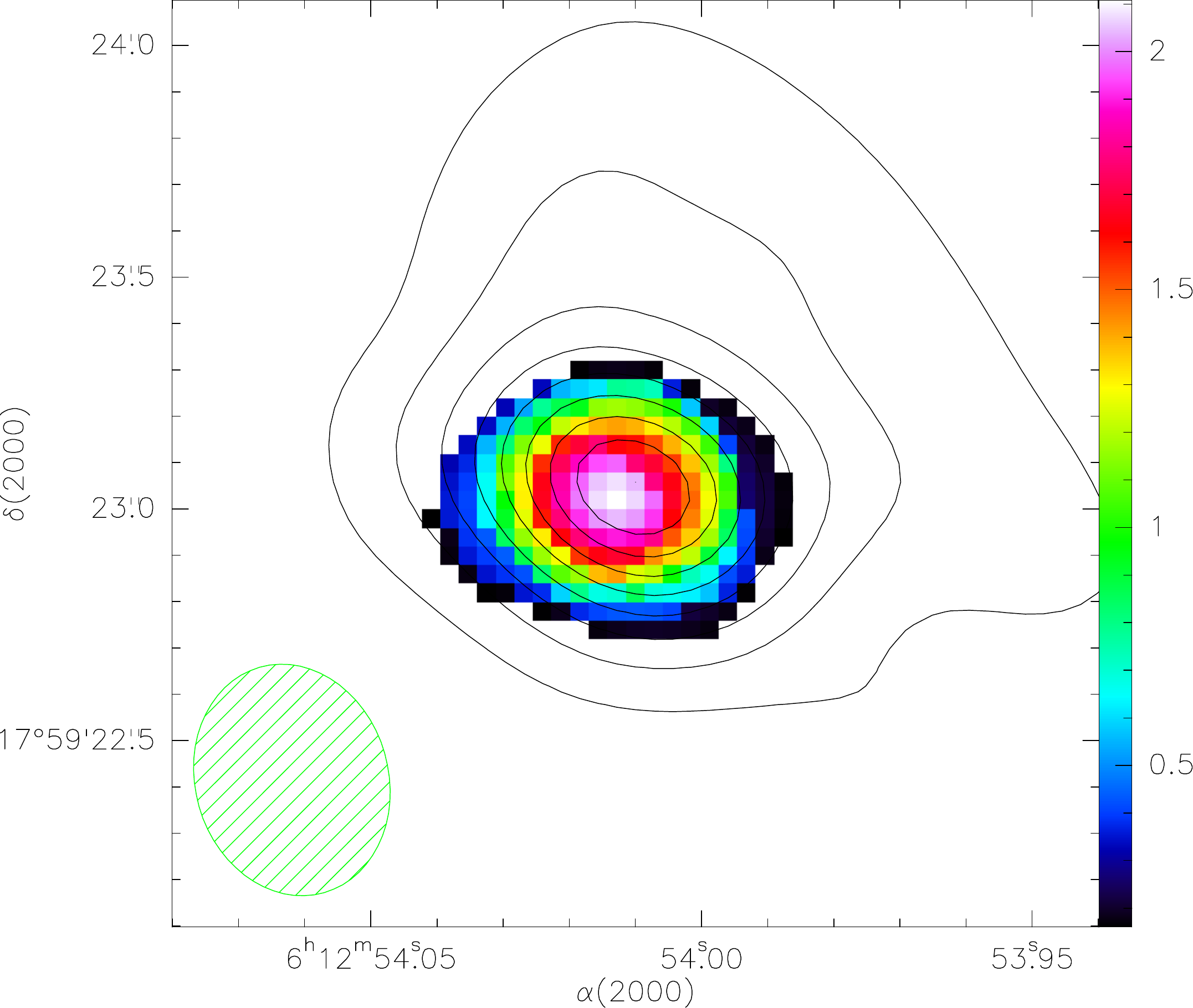}
\end{minipage}
\hfill
\begin{minipage}{0.49\textwidth}
\centering
    \includegraphics[width=\columnwidth]{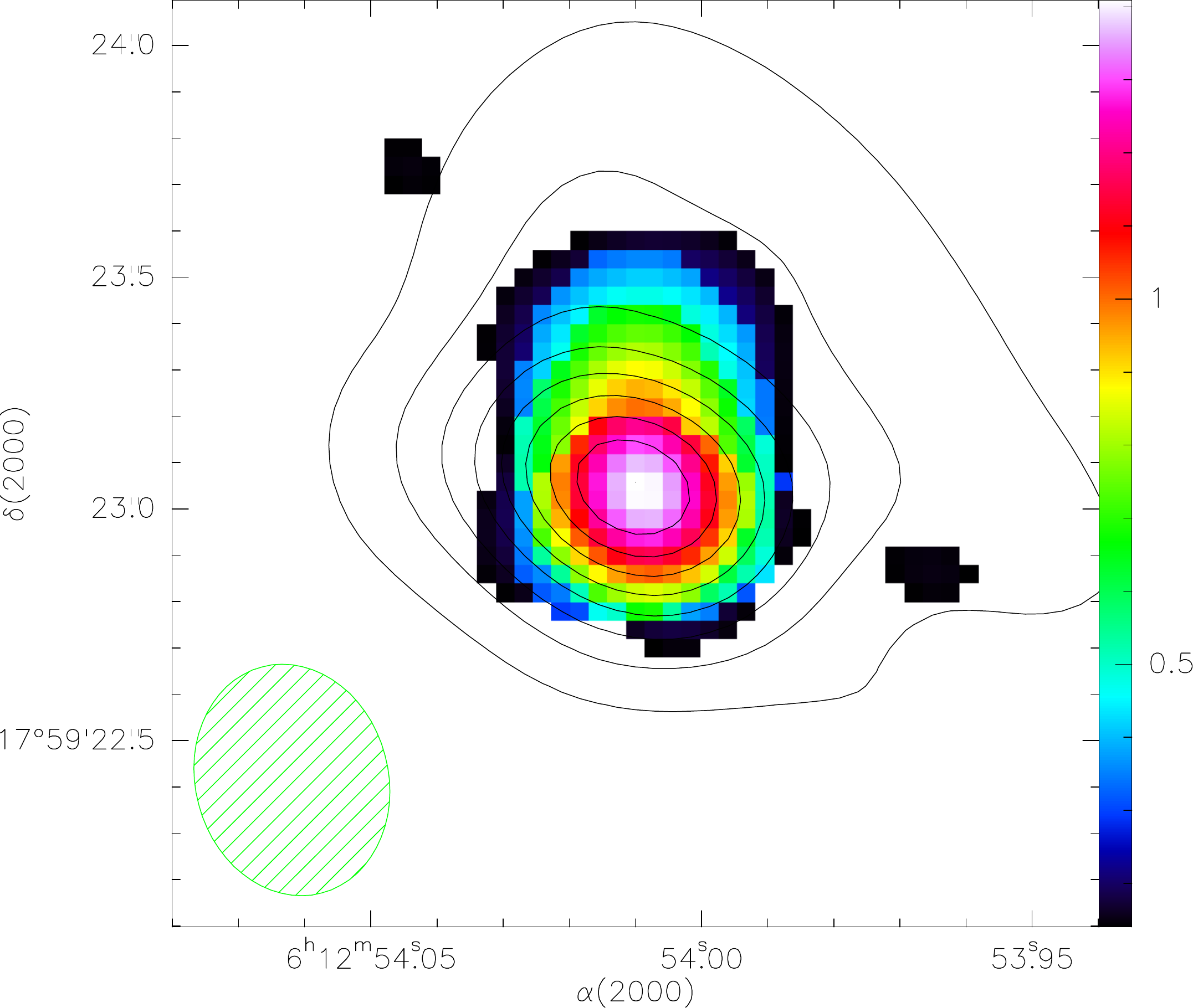}
\end{minipage}
  \caption{The maps of the integrated intensity of HNCO $ J_{\mathrm{K}_{-1}} = 10_{2}-9_{2} $ (left panel) and OCS $ J=18-17 $ (right panel) transitions (color scale) towards the SMA1 clump. The intensity units are Jy\,beam$^{-1}$\,km\,s$^{-1}$. The beam size is 0$\farcs$4. The contours show the 1.3~mm continuum emission with the same angular resolution.  {The beam for the molecular maps is shown in the lower left corner of both panels.}}
\label{fig:sma1_hnco}
\end{figure*}

This diagram indicates a rotational temperature of $318\pm 70$~K (Table~\ref{table:rd}).  {However, the $ K_{-1} = 0, 1 $ transitions can be saturated. The peak brightness temperature in these transitions is close to the brightness temperature in the apparently optically thick lines (e.g. CO and CH$_3$CN) and our modeling using RADEX \citep{vdTak07} shows that at the derived physical parameters and column density the optical depth in these lines is about unity. In this case the derived rotational temperature would apparently represent an upper limit to the excitation temperature.} The deconvolved source size is $ \sim 0\farcs3 \times 0\farcs2 $ for the $ K_{-1} = 0 $ transition and decreases for higher $K_{-1}$ ladders. The map of the integrated intensity in the HNCO $ K_{-1} = 2 $ transition is shown in Fig.~\ref{fig:sma1_hnco} along with the map of the OCS emission which is another tracer of hot gas. The peaks of the HNCO and OCS emission practically coincide with the continuum peak, although distributions of these molecules seem to be somewhat different.

\begin{deluxetable}{lccc}
\tablecaption{Fit results for the rotational diagrams obtained towards SMA1. The 1$\sigma$ uncertainties are indicated in the parentheses. In the last column the approximate HPBW for the corresponding data is indicated. \label{table:rd}}
\tablehead{\colhead{Molecule} &\colhead{$T_\mathrm{kin}$} &\colhead{$ \log N_\mathrm{L} $} &\colhead{HPBW} 
\\ 
\colhead{ } &\colhead{(K)} &\colhead{(cm$^{-2}$)} &\colhead{($ ^{\prime\prime} $)}
}
\tablecolumns{4}
\startdata
HNCO & 318(70)  &16.34(0.04) &$ \approx $0.4\\
SO$_2$ & 146(16) &15.50(0.05) &$ \approx $2 \\
             & 65(11)    &15.35(0.08) &$ \approx $2 
\enddata
\end{deluxetable}

The total HNCO column density derived from the rotational diagram is $ \sim 2\times 10^{16} $~cm$^{-2}$. We used the partition function from the Cologne Database for Molecular Spectroscopy \citep{Mueller01,Mueller05}, too. A comparison with the total gas column density indicates HNCO abundance of $ X(\mathrm{HNCO}) \sim 10^{-8} $. This is a rather high value, close to the highest HNCO abundance derived in the survey of massive cores by \citet{Zin00}.  {However, taking into account the note above about the HNCO lines saturation, this value may need a correction.}

\begin{figure}
\centering
\includegraphics[width=\columnwidth]{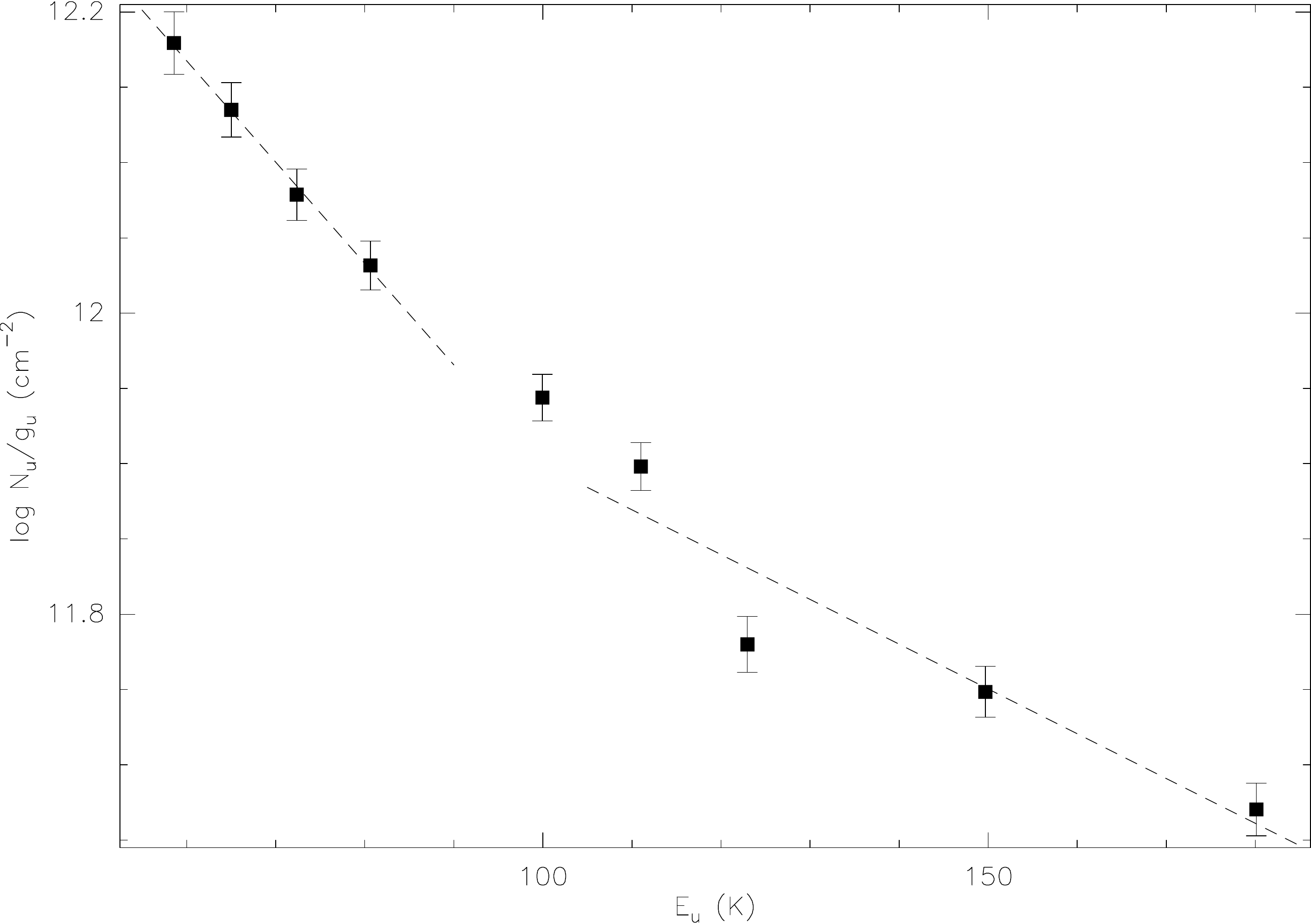}
\caption{The population diagram for SO$_2$ at the SMA1 emission peak.}
\label{fig:so2-rd}
\end{figure}

In the 350~GHz band we detected many SO$_2$ lines (Table~\ref{table:lines-comp}). The corresponding population diagram is shown in Fig.~\ref{fig:so2-rd}. It indicates a range of temperatures. For low-excitation transitions the rotational temperature is $65\pm  11$~K. For the high-excitation ones it is $146\pm 16$~K. The SO$_2$ observations were performed at a much lower angular resolution (2 arcsec) than in case of HNCO and apparently include emission from both the hot core and the surrounding cloud. The derived temperatures are consistent with the other estimates for these components.

Another indicator of the hot environment is the vibrationally excited HCN. We detected the emission in the $ v_2 =1 $ state, about 1000~K above the ground state. For the $ v_2 =2 $ emission the upper limit is about 5 times lower. Following the analysis presented in e.g.  \citet{Veach13} we obtain an upper limit for the excitation temperature between these states of about 500~K. This is consistent with the other estimates of the gas temperature given above. The critical density for HCN excitation is high ($ > 10^{10} $~cm$^{-3}$, \citealt{Veach13}) which is consistent with our density estimates for the hot gas.

\subsubsection{A cold clump in the hot core?}

\begin{figure*}
\begin{minipage}{0.49\textwidth}
\centering
    \includegraphics[width=\columnwidth]{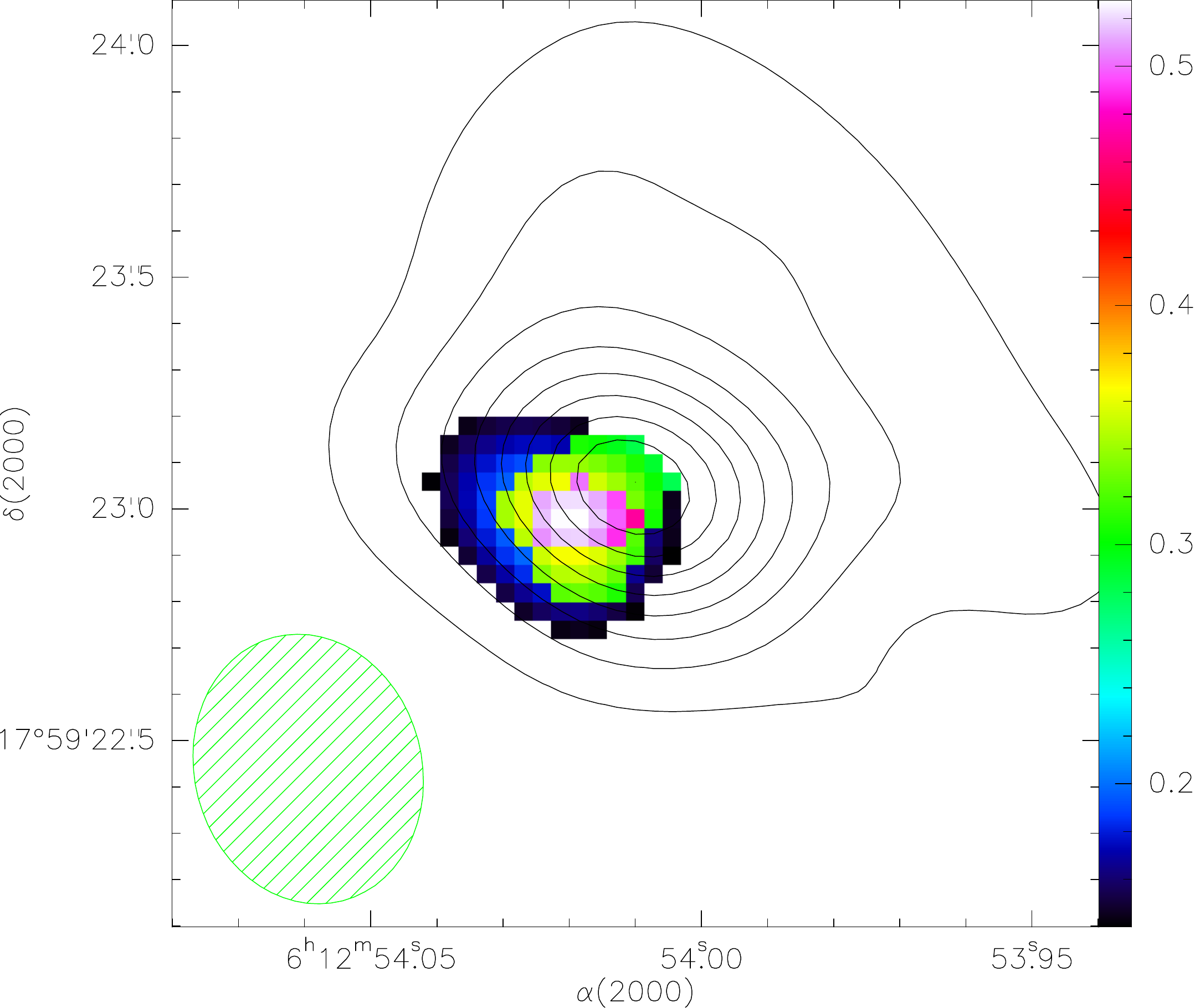}
\end{minipage}
\hfill
\begin{minipage}{0.49\textwidth}
\centering
    \includegraphics[width=\columnwidth]{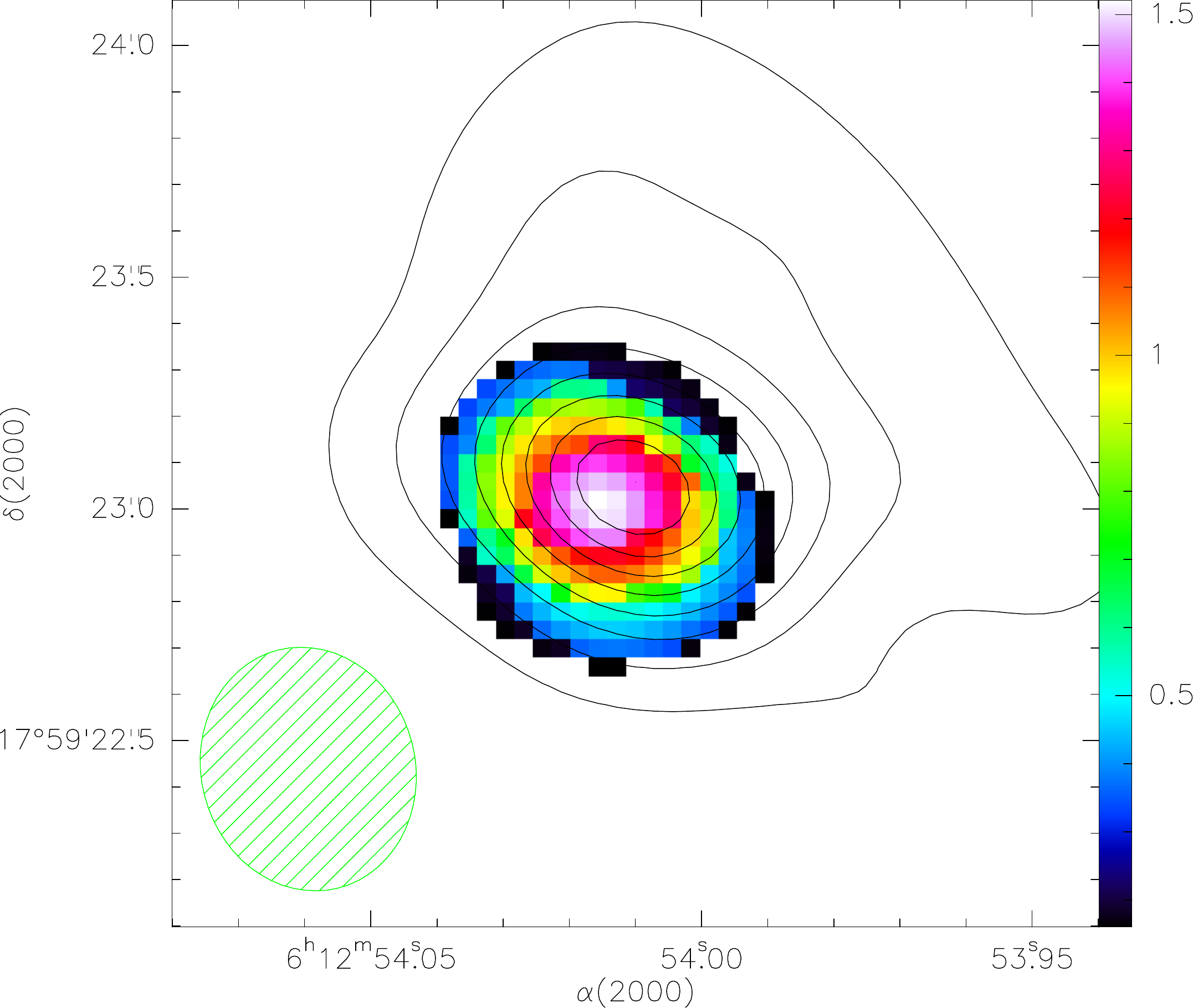}
\end{minipage}
  \caption{The maps of the integrated intensity of the DCN $J=3-2$ (left panel) and $^{13}$CS $J=5-4$ (right panel) transitions (color scale) towards the SMA1 clump. The intensity units are Jy\,beam$^{-1}$\,km\,s$^{-1}$. The beam size is 0$\farcs$4. The contours show the 1.3~mm continuum emission with the same angular resolution.  {The beam for the molecular maps is shown in the lower left corner of both panels.}}
\label{fig:sma1_dcn}
\end{figure*}

In Fig.~\ref{fig:sma1_dcn} we plot maps of the DCN $J=3-2$ and $^{13}$CS $J=5-4$ integrated line emission in the SMA1 core. The emission regions are very compact and the emission peak is shifted from the continuum peak which apparently coincides with the YSO location. The projected distance from the continuum peak is roughly 300~AU. 

The velocities of the DCN and $^{13}$CS emission are practically the same as those of high excitation lines of other molecules tracing apparently the hot gas. Therefore, most probably the observed DCN emission arises within the disk and since we probably see the disk nearly face-on as discussed above, the physical distance from the center is not much larger than the projected distance. An estimate of the abundances in the LTE approximation gives $ X(\mathrm{DCN}) > 10^{-11} $ and $ X(\mathrm{^{13}CS}) > 3\times 10^{-11} $ (assuming the total mass of the clump $ <10 $~M$_\odot$ and temperature of 50--100~K). This is a ``normal'' value for $^{13}$CS while the derived DCN abundance implies a significant deuteriation as follows from comparison with typical HCN abundances in massive cores \citep[e.g.][]{Zin09}. As pointed out in Paper~I, recent modelling by \citet{Albertsson11} shows that DCN/HCN abundance ratio sharply drops at temperatures $ \ga 80 $~K. It means that temperature of the DCN emitting clump should be rather low, much lower than the temperature of the hot gas in the disk. Another possible explanation, as also discussed in Paper~I, can be a very young age of the clump, insufficient to reach the steady-state DCN/HCN abundance ratio, but this looks less probable. In principle, interiors of the accretion disk can be well shielded from an external radiation, probably providing necessary low temperatures. In any case this clump deserves further investigation in order to clarify its properties and nature. It is worth mentioning that the mass of this clump cannot be lower than $\sim 1 $~M$_\odot$, otherwise the inferred $^{13}$CS abundance would be unrealistically high. The virial mass we estimate about the same, hence the clump can be gravitationally bound.

\subsection{SMA2} \label{sec:sma2}
As in case of the SMA1 the spectral slope for the continuum emission in the frequency range 284--350~GHz is too low ($ \sim 1.3 $), which is probably caused by a higher flux loss at higher frequencies. With the SMA at sub-arcsecond resolution we detected a compact component in continuum emission of about the same size as in the SMA1. However, no high-excitation molecular lines could be detected in this area. In Paper~I we derived the temperature of the SMA2 clump $ \sim 40 $~K. Now with the extended data set, from the methanol excitation analysis (Sect.~\ref{sec:methanol_analysis})  we found the 1$\sigma$ confidence interval for temperature of 40--80~K. The best estimate of the methanol relative abundance is $ X(\mathrm{CH_3OH})\sim 10^{-8} $. These estimates are obtained at the 2 arcsec scale. It cannot be excluded that the temperature of the compact structure observed in the very extended configuration is somewhat higher. However it cannot be much higher since there is no sign of a higher temperature component in the molecular data. Assuming the temperature of 50~K we obtain the mass of the compact core $\sim 0.2$~M$_\odot$.

To the west from the SMA2 there is an area of molecular emission without continuum counterpart in the SMA data, observed earlier in the N$_2$H$^+$, NH$_3$ and several CH$_3$OH lines (Paper~I and \citealt{Wang11}). Our new methanol data analysis indicates the temperature in the range 25--65~K, density $ n(\mathrm{H_2}) $ in the range $6.3 \times10^{4}$---$2.5 \times10^{6}$, $ X(\mathrm{CH_3OH})\sim 10^{-7}$. Some of the methanol transitions are inverted in the model and can be masing.
These high gas densities and rather high methanol abundances at relatively low temperatures can be explained by the influence of shock.

\begin{deluxetable*}{cccccc}
\tablecaption{Physical parameters derived from the CH$_3$OH data obtained in the compact configuration (HPBW $ \approx 3^{\prime\prime} $). The 1$\sigma$ confidence intervals are indicated in the parentheses. \label{table:ch3oh-sma2-4}}
\tablehead{\colhead{Object} &\colhead{$T_\mathrm{kin}$} &\colhead{$N_{\rm{CH_3OH}}/{\rm\Delta}V  $} &\colhead{$n_{\rm H_2}$} &\colhead{beam filling factor} &\colhead{$N_{\rm{CH_3OH}}/N_{\rm H_2}$}\\ 
\colhead{ } &\colhead{(K)} &\colhead{($10^{12}$ cm$^{-3}$s)} &\colhead{(cm$^{-3}$)} &\colhead{(\%)} &\colhead{ }
}
\tablecolumns{6}
\startdata
SMA2             & 50 (40---80)&	 2.0 (1.8---3.5)  &   6.3$\times 10^{6}$ ($\leq10^{8}$)                  &  99.3 ($\geq 84$) &  $10^{-8} (\geq10^{-8})$  \\
SMA2-W         & 40 (25---65)&	 3.2 (2.2---4.5)  &   1.0$\times 10^{6}$ ($6.3 \times10^{4}$---$2.5 \times10^{6}$)   &  64.0 (63---70) &  $10^{-7} (\geq10^{-8})$\\ 
SMA4              & 45 (30---90)&	 3.6 (0.8---7.9)  &   5.6$\times 10^{4}$ ($10^{3}$---$6.3\times 10^{5}$) &  99.9 ($\geq 20$) &  $10^{-7} (10^{-6}$---$10^{-9})$ 
\enddata

\end{deluxetable*}
\subsection{SMA3} \label{sec:sma3}
In Paper~I we did not detect this component which coincides with the near-infrared source NIRS~1 \citep{Tamura91} and is identified as a massive disk candidate by NIR polarization observations \citep{Jiang08}. However, \citet{Simpson09} argued that these measurements could be affected by instrumental effects and there is no real evidence for the ``polarization disk'' here. At the same time the NIR measurements indicate a possible outflow related to this object. A comparison with the measurements by \citet{Wang11} gives a spectral index of about 3.3 in the frequency range 225--350~GHz, consistent with the optically thin dust emission.    \citet{Wang11} estimated mass of this clump at about 2~M$_\odot$. At sub-arcsecond resolution we do not see any continuum (Fig.~\ref{fig:cont}) or molecular emission at this position. 

\subsection{SMA4} \label{sec:sma4}
This clump was first detected in Paper~I. It shows a weak continuum and spectral line emission in several molecular transitions. The measured continuum flux density at 350~GHz is practically the same as at 284~GHz which also implies a much higher flux loss at the higher frequency. The temperature derived from our new methanol data analysis is about 45~K, higher than estimated in Paper~I from ammonia data. However the ammonia emission here is very weak and the uncertainties are high. Some methanol transitions are inverted in the model. The influence of shock is also probable.

\section{Morphology and properties of the outflows}

\begin{figure*}
\centering
\includegraphics[angle=-90,width=\textwidth]{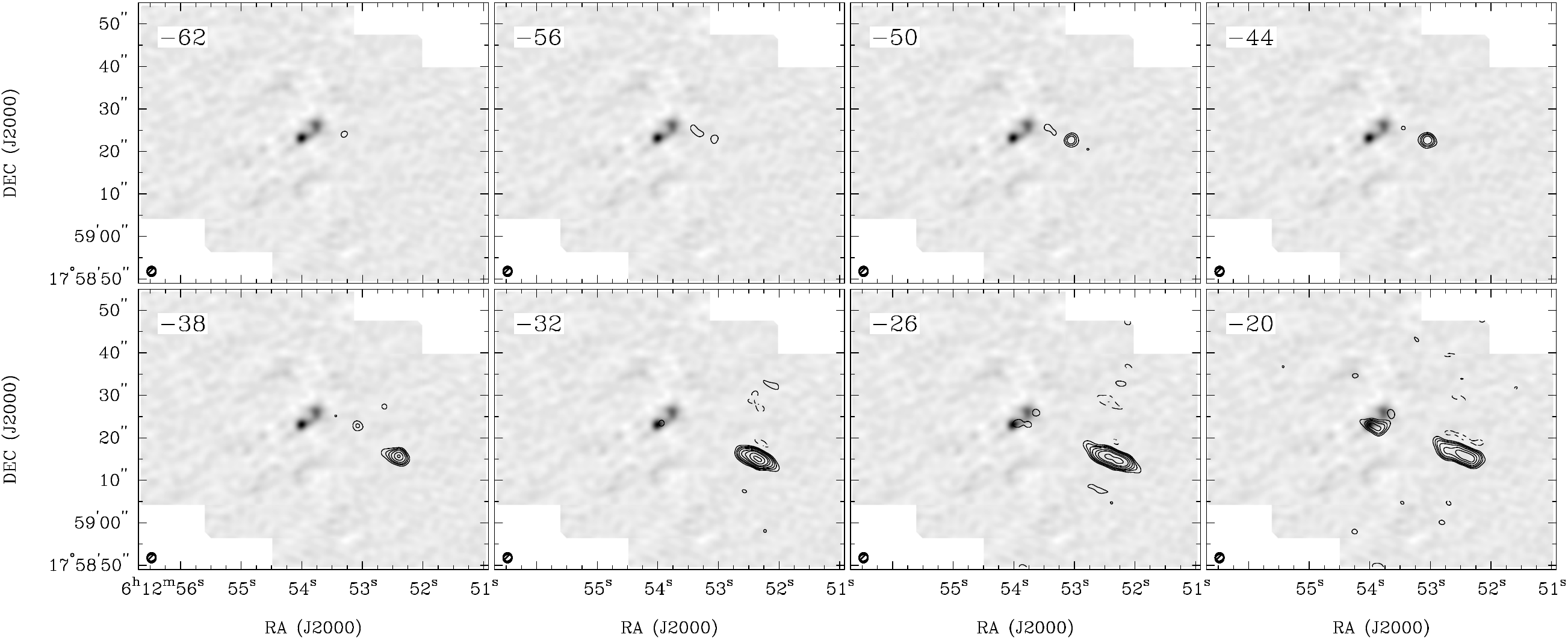}
\caption{Channel maps of the CO blue wing emission in the S255IR. The numbers in the upper left corner indicate the channel velocity in km\,s$^{-1}$. The dashed contours show negative features due to the missing flux. The contour levels are (--3, 3, 5, 7, 10, 15, 20, 30, 40)$\times 100$~mJy\,beam$^{-1}$. The SMA beam is shown in the lower left corner of each panel.}
\label{fig:chmap-co-blue}
\end{figure*}

\begin{figure*}
\centering
\includegraphics[angle=-90,width=\textwidth]{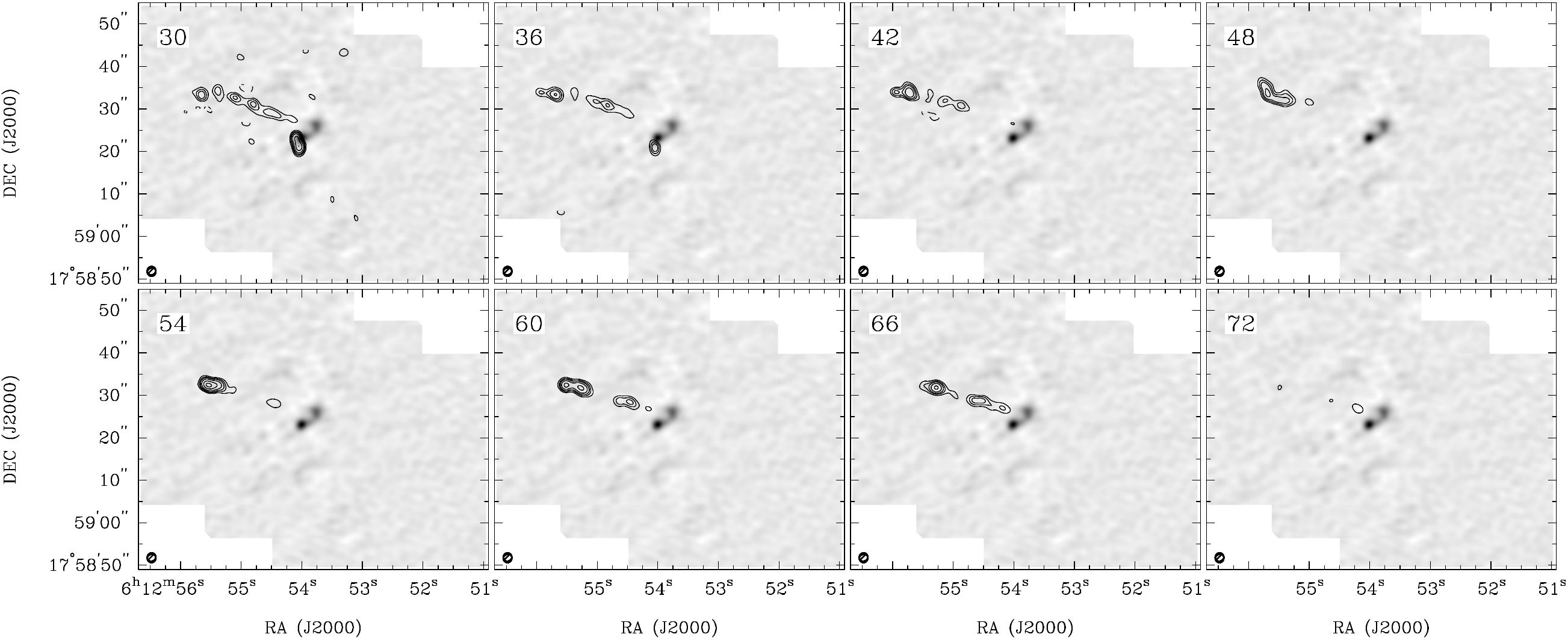}
\caption{Channel maps of the CO red wing emission in the S255IR. The numbers in the upper left corner indicate the channel velocity in km\,s$^{-1}$. The dashed contours show negative features due to the missing flux. The contour levels are (--3, 3, 5, 7, 10, 15, 20, 30, 40)$\times 100$~mJy\,beam$^{-1}$. The SMA beam is shown in the lower left corner of each panel.}
\label{fig:chmap-co-red}
\end{figure*}

\begin{figure*}[htb]
\begin{minipage}{0.49\textwidth}
\centering
\includegraphics[angle=-90,width=\columnwidth]{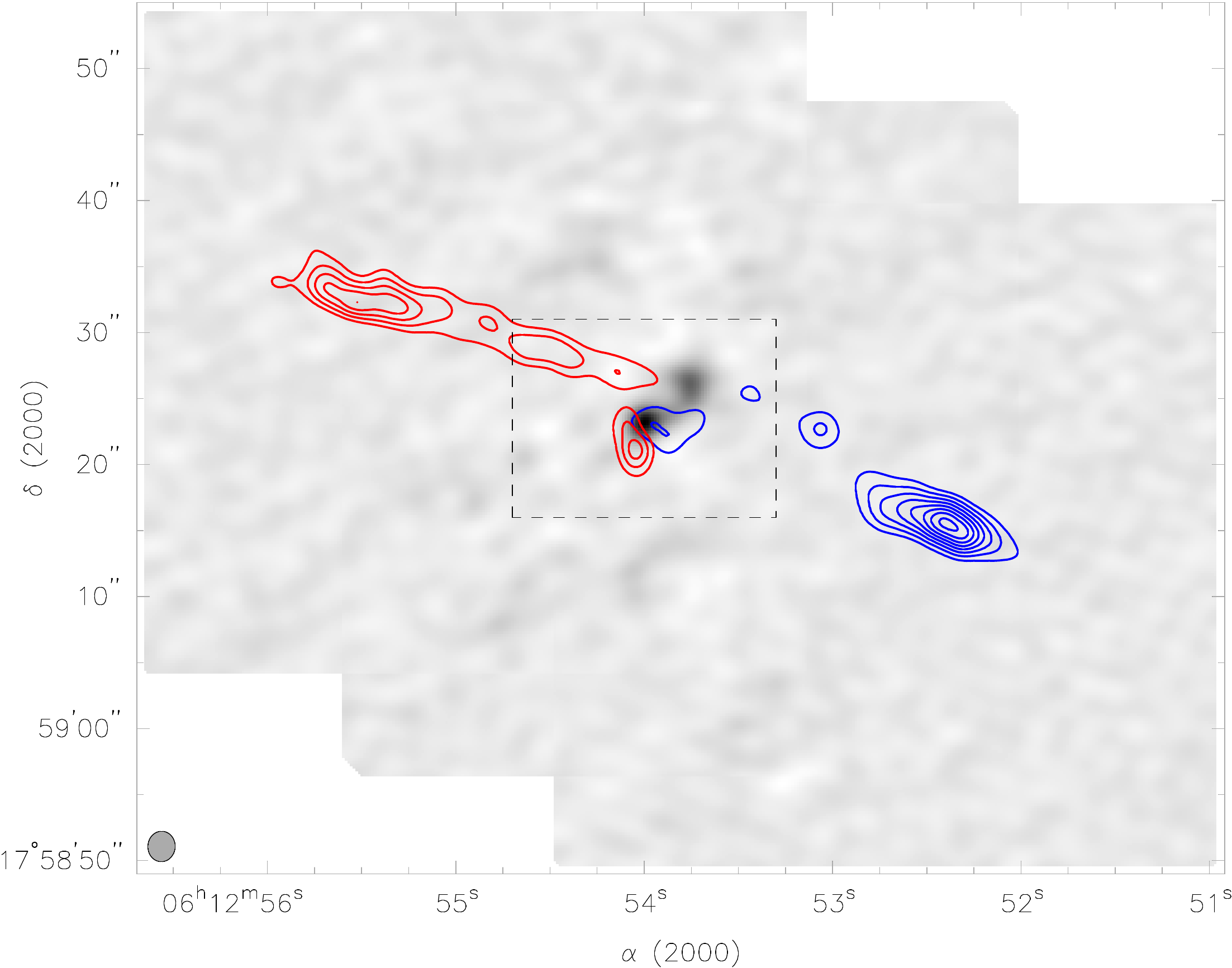} 
\end{minipage}
\hfill
\begin{minipage}{0.49\textwidth}
\centering
\includegraphics[angle=-90,width=\columnwidth]{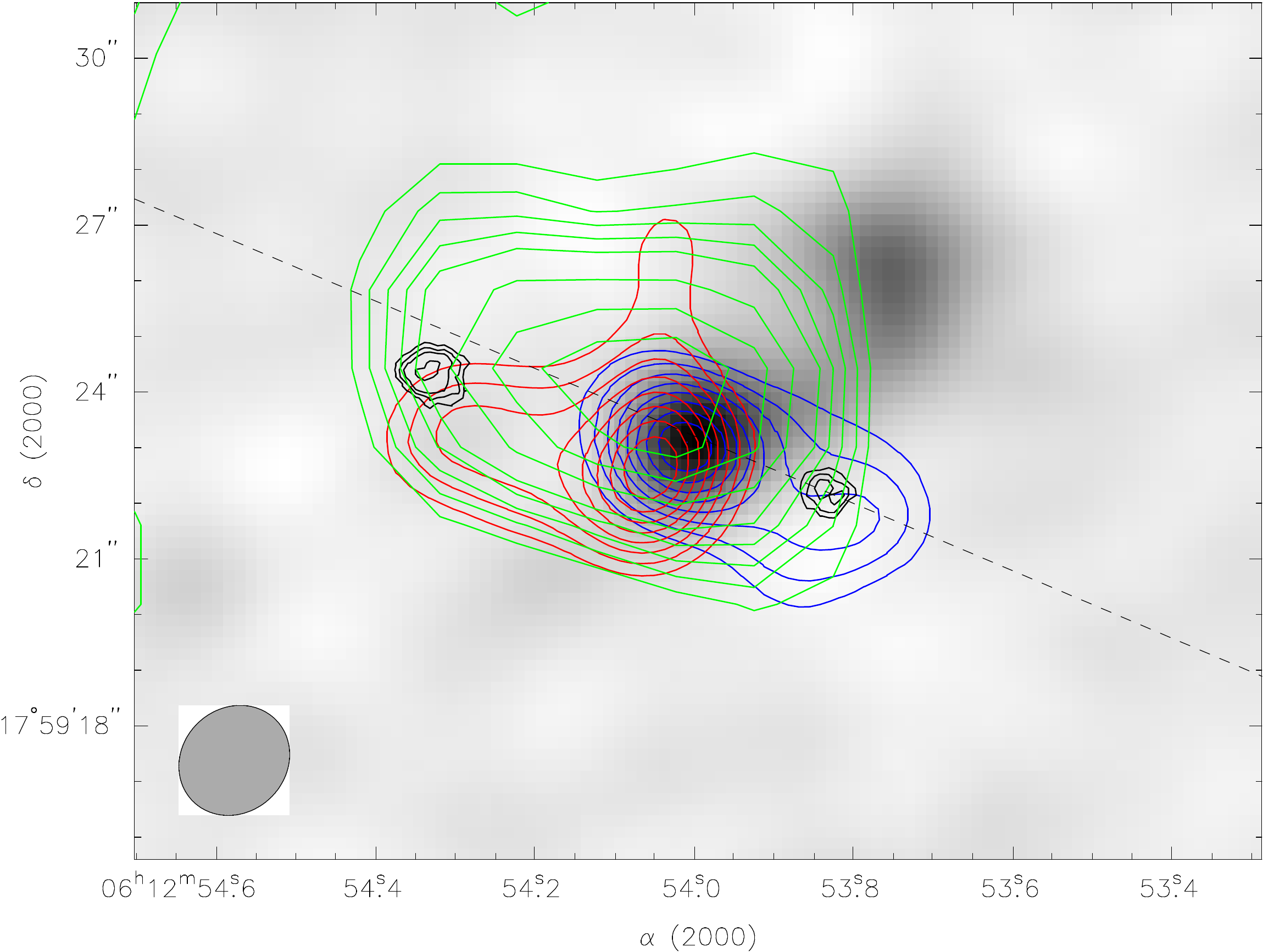}
\end{minipage}
\caption{Left panel: maps of the CO(3--2) high velocity emission as observed with the SMA (blue and red contours) in the S255IR area overlaid on the continuum image at 0.8~mm. The SMA1 and SMA2 continuum clumps are marked. The velocity intervals are from --63~km\,s$^{-1}$ to --13~km\,s$^{-1}$ for the blue wing and from 25~km\,s$^{-1}$ to 75~km\,s$^{-1}$ for the red wing. The dashed rectangle indicates the area shown in the right panel. Right panel: maps of the HCO$^+$(4--3) high velocity emission (blue and red contours) in the S255IR area overlaid on the continuum image at 0.8~mm. The velocity intervals are from --19~km\,s$^{-1}$ to 3~km\,s$^{-1}$ for the blue wing and from 11~km\,s$^{-1}$ to 27~km\,s$^{-1}$ for the red wing. The black contours show the \Fetwo\ emission \citep{Wang11}. The map of the 15~GHz continuum emission (from the VLA archival data, the angular resolution is about 4 arcsec) is plotted with green contours. The dashed line indicates the jet axis (PA = 67$^\circ$) as in Fig.~\ref{fig:sma1_ch3oh-vel}.  {The SMA beam is shown in the lower left corner of both panels.}}\label{fig:co32-outflow} 
\end{figure*}

In the S255IR area high velocity emission is detected in lines of CO, SiO and several high density tracers including HCN, HCO$^+$ and CS. In Figs.~\ref{fig:chmap-co-blue},\ref{fig:chmap-co-red} we present channel maps of the CO $ J=3-2 $ emission in the blue and red line wings, respectively. The maps of the integrated line wing emission (Fig.~\ref{fig:co32-outflow}) show that the CO high velocity emission observed with the SMA looks like a highly collimated bipolar outflow originating near SMA2. There is another more compact component near SMA1. 

On the other hand the bipolar outflow observed in HCN and HCO$^+$ is apparently associated with SMA1 while is parallel to the CO flow (Fig.~\ref{fig:co32-outflow}). It is apparently associated with the jet observed in particular in the \Fetwo\ emission which is also plotted in this figure. It is worth mentioning that the extent of the jet is much smaller than the extent of the outflow seen in CO, although their orientations coincide.

\begin{figure}
\centering
\includegraphics[angle=-90,width=\columnwidth]{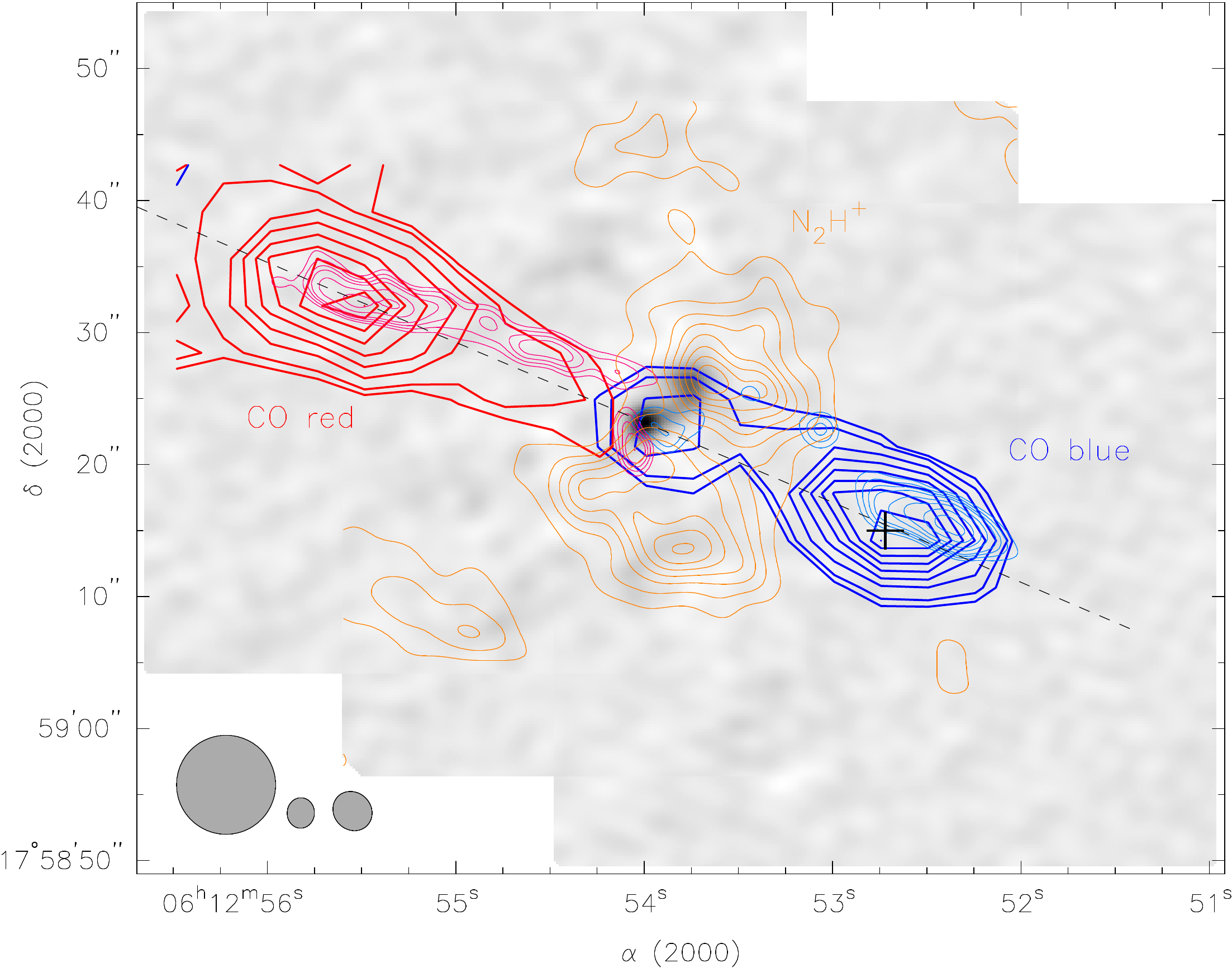}
\caption{Maps of the CO(3--2) high velocity emission as observed with the IRAM 30m telescope (blue and red thick contours) in the S255IR area overlaid on the continuum image at 0.8~mm. The thin contours show the SMA maps from Fig.~\ref{fig:co32-outflow}. The dashed line indicates the axis of the position-velocity cut (PA = 67$^\circ$). The orange contours show the N$_2$H$^+$(3--2) integrated line emission obtained by combining the SMA and 30m data. The cross marks the position of the high velocity dense clump (Sect.~\ref{sec:hvc}).  {The 30m beam and the SMA beams for the CO and N$_2$H$^+$ observations, respectively, are shown in the lower left corner (from left to right).}}
\label{fig:CO-30m}
\end{figure}

The SMA data alone hint at two parallel outflows with different origins. However single-dish CO(3--2) observations with the IRAM 30m telescope show a different picture. Here we see a wider outflow clearly originating at SMA1 (Fig.~\ref{fig:CO-30m}).  {While the emission peaks} coincide in the SMA and 30m maps, near the driving source the CO emission measured with the SMA traces only the northern edge of the outflow. This shows that the SMA map gives a distorted picture and in reality we have here a wider angle outflow originating at the SMA1. 

Apparently this means that this northern edge of the outflow contains a relatively bright component with a characteristic scale comparable to the SMA beam. The  {southern edge} of the outflow near the driving source  {would be} more diffuse and resolved out by the SMA. Most probably this implies that the CO emission is formed in a compressed layer surrounding the outflow cavity. For some (unclear) reason in the northern part this layer is more pronounced. It is worth mentioning that the brightest peaks of the high velocity CO emission practically coincide in the SMA and 30m maps and also coincide with spots of molecular hydrogen emission \citep{Wang11} which probably indicate bow shocks. Basic physical parameters of the outflow were estimated by \citet{Wang11} from their CO(2--1) data. Our new data are consistent with these estimates.

\begin{figure*}
\begin{minipage}{0.49\textwidth}
\centering
\includegraphics[angle=-90,width=\columnwidth]{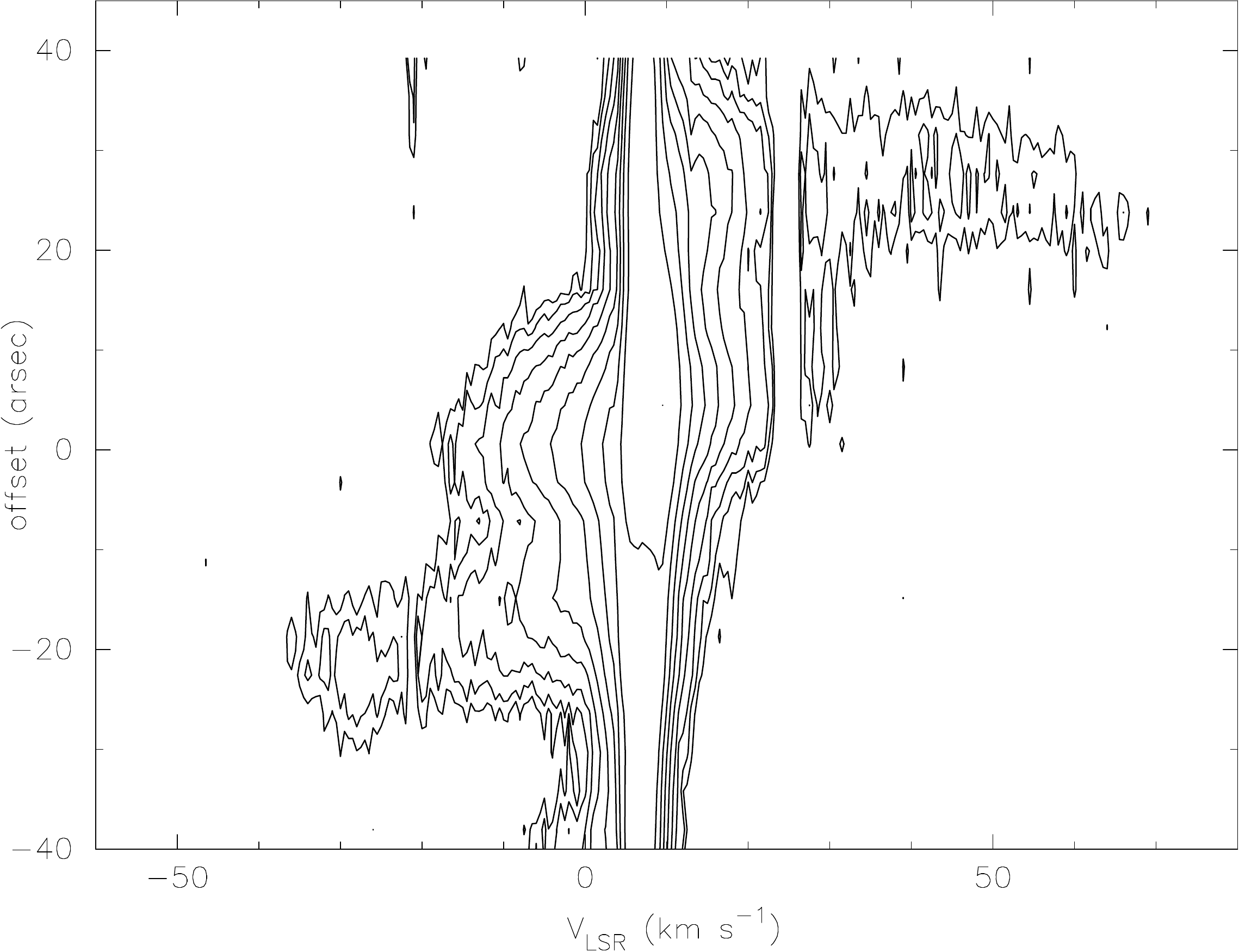}
\end{minipage}
\hfill
\begin{minipage}{0.49\textwidth}
\centering
\includegraphics[angle=-90,width=\columnwidth]{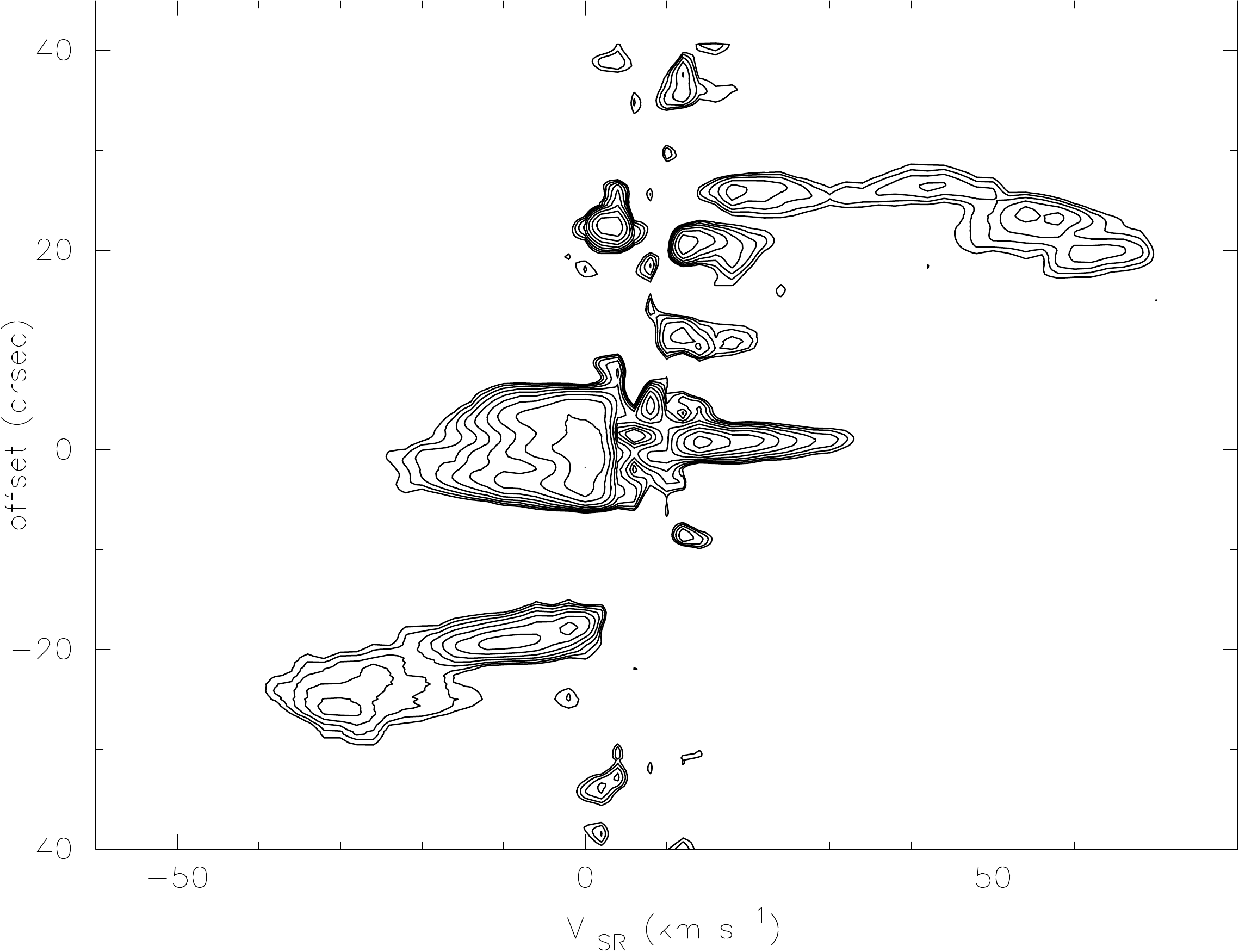}
\end{minipage}
\caption{Left panel: the position-velocity map for the CO(3--2) emission as observed with the IRAM 30m telescope along the cut indicated in Fig.~\ref{fig:CO-30m}. The contour levels are on logarithmic scale  from 0.4~K to 31.5~K. The gap at $V_\mathrm{LSR} \sim 24$~km\,s$^{-1}$ is caused by the emission at the reference offset position mentioned above. Right panel: the position-velocity map for the CO(3--2) emission as observed with the SMA along the same cut. The contour levels are on logarithmic scale  from 0.5 to 11.5~Jy\,beam$^{-1}$.}
\label{fig:CO-pv}
\end{figure*}

The position-velocity (P--V) diagram for the outflow along the cut indicated in Fig.~\ref{fig:CO-30m}, constructed from the 30m data, is shown in the left panel of Fig.~\ref{fig:CO-pv}. The gap at $V_\mathrm{LSR} \sim 24$~km\,s$^{-1}$ is caused by the emission at the reference offset position mentioned above.  {In the right panel we present the P--V diagram obtained from the SMA CO(3--2) data. The main features of these diagrams coincide. They show non-monotonic dependence of the velocity on offset, which hints at two outflow components at different distances from the driving source. These P--V diagrams are somewhat different from that presented by \citet{Wang11}. However \citet{Wang11} plotted this diagram for a different cut, with the position angle of 75$^\circ$, which makes the direct comparison irrelevant.}

\begin{figure*}
\begin{minipage}{0.49\textwidth}
\centering
\includegraphics[angle=-90,width=\columnwidth]{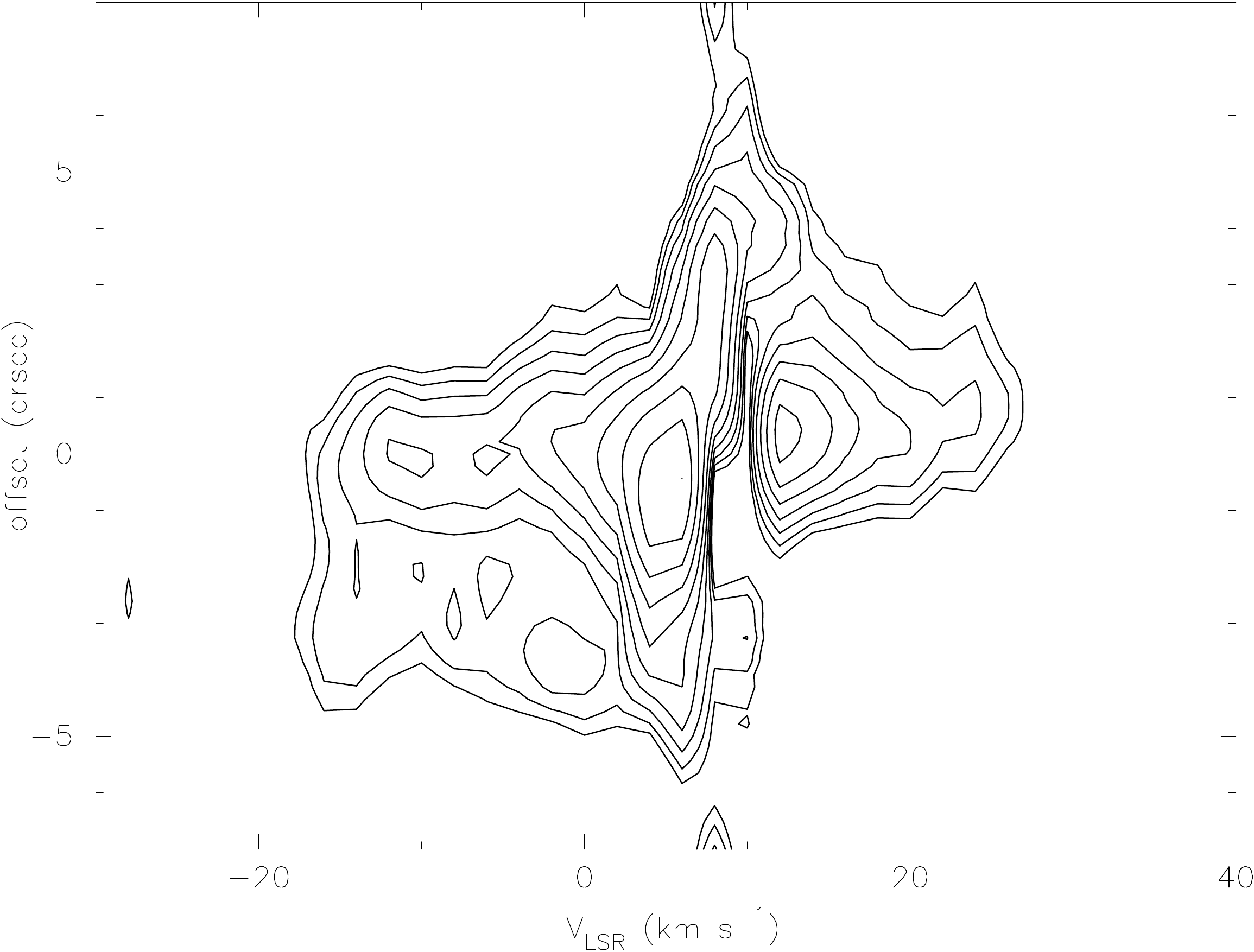}
\end{minipage}
\hfill
\begin{minipage}{0.49\textwidth}
\centering
\includegraphics[angle=-90,width=\columnwidth]{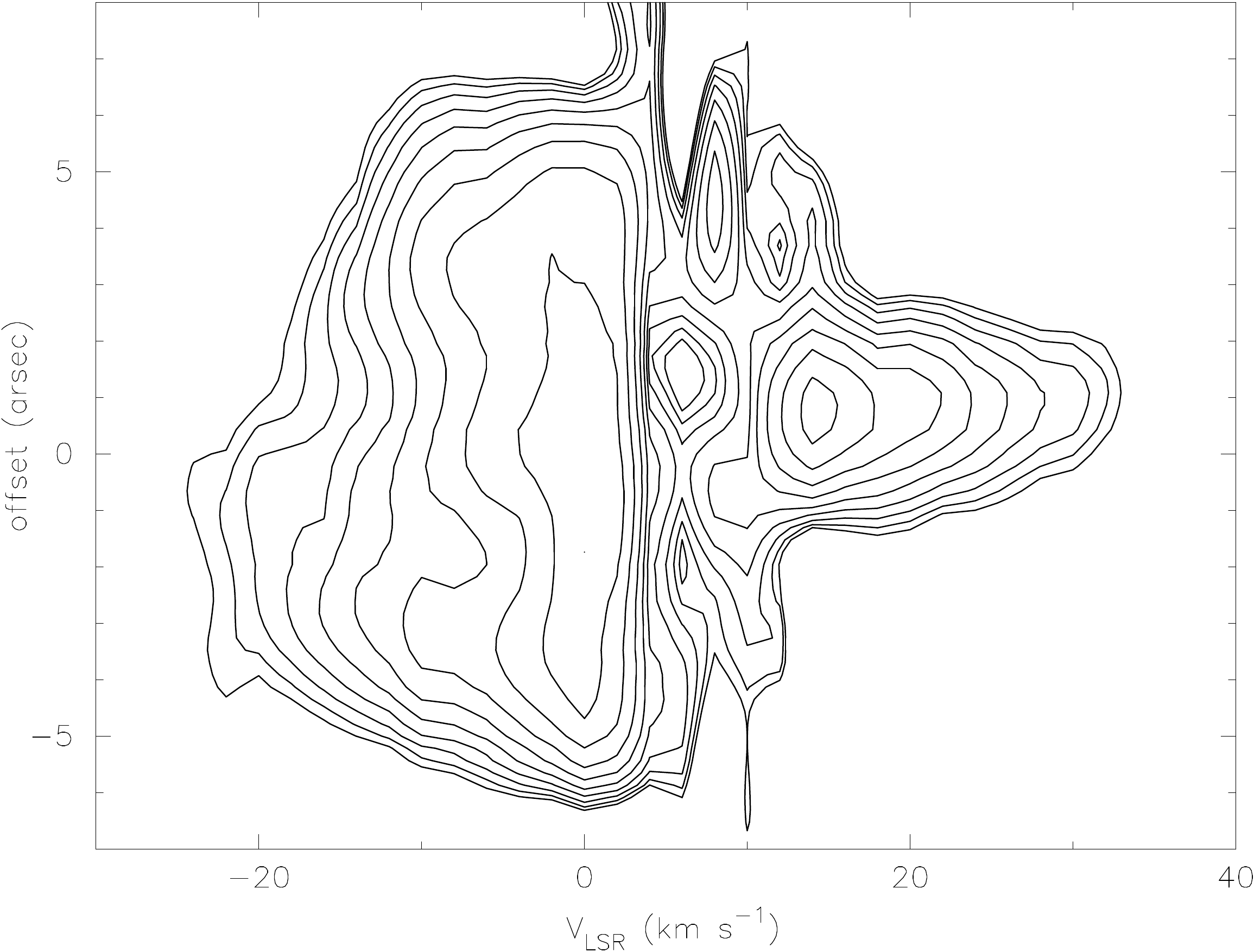}
\end{minipage}
\caption{Left panel: the position-velocity map for the HCO$^+$(4--3) emission as observed with the SMA along the cut indicated in Fig.~\ref{fig:co32-outflow}. The contour levels are on logarithmic scale  from 0.4 to 7.6~Jy\,beam$^{-1}$. Right panel: a part of the position-velocity map for the CO(3--2) emission measured by the SMA from Fig.~\ref{fig:CO-pv} for the same intervals of the offset and velocity as in the left panel.}
\label{fig:HCO-pv}
\end{figure*}

{In Fig.~\ref{fig:HCO-pv} we plot the P--V diagram for the HCO$^+$(4--3) emission as observed with the SMA, along with the part of the CO(3--2) P--V diagram for the same intervals of the offset and velocity. The diagrams are very similar, which means that both molecules trace apparently the same gas.} 

\subsection{Dense high velocity clump} \label{sec:hvc}
\begin{figure}
\includegraphics[width=\columnwidth]{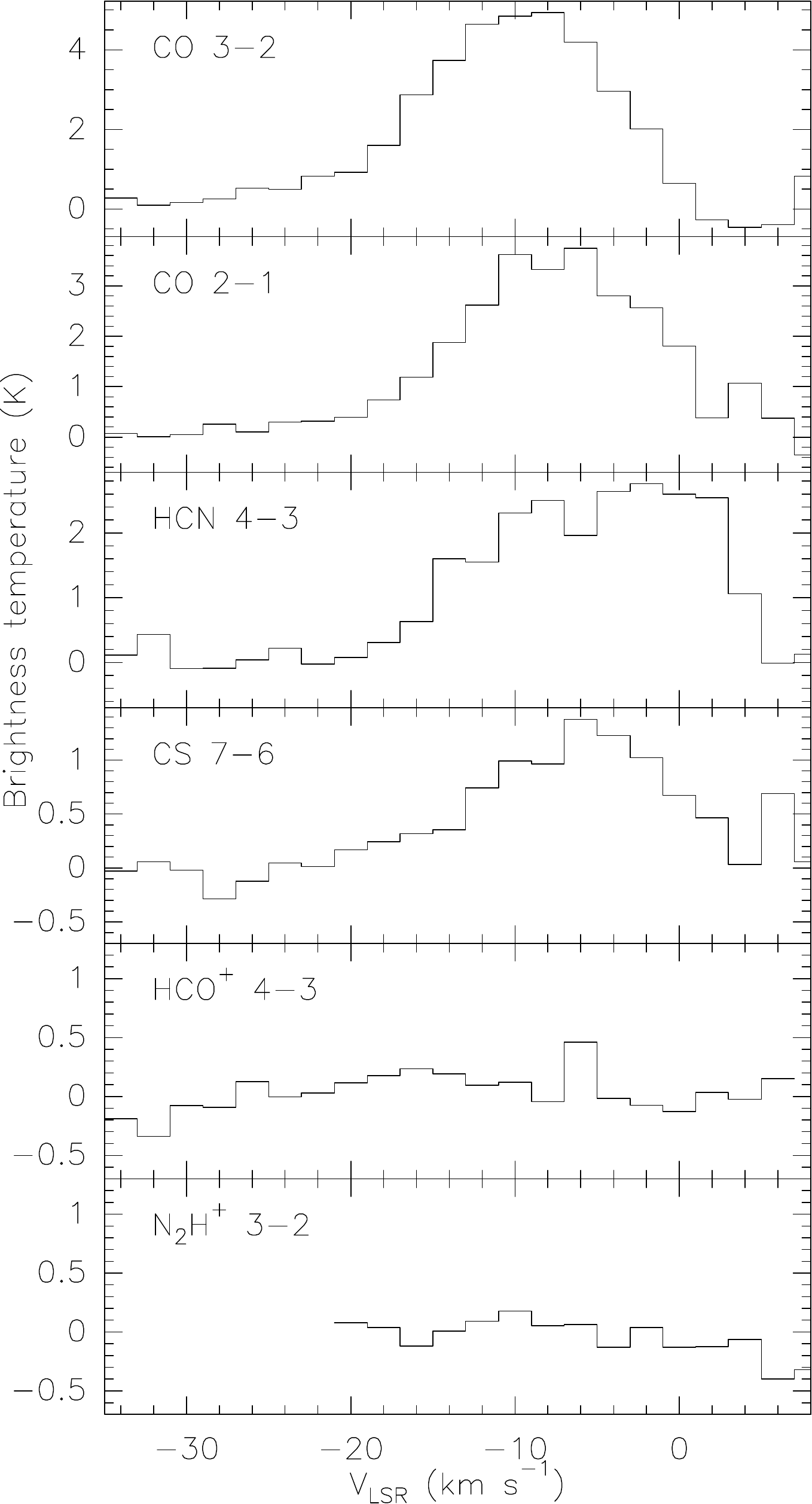}
\caption{The CO(3--2), CO(2--1), HCN(4--3), CS(7--6), HCO$^+$(4--3) and N$_2$H$^+$(3--2) spectra towards the position marked by cross in Fig.~\ref{fig:CO-30m}.}
\label{fig:hvc-spectra}
\end{figure}

Our data show a {strong, compact} blue-shifted CS and HCN emission close to the peak of the CO blue-shifted line wing emission. The peak of this CS and HCN emission is marked by cross in Fig.~\ref{fig:CO-30m}. The CO(3--2), CO(2--1), HCN(4--3), CS(7--6), HCO$^+$(4--3) and N$_2$H$^+$(3--2) spectra towards this position are presented in Fig.~\ref{fig:hvc-spectra}.

One can see rather strong and broad HCN and CS lines at a central velocity of about --5~km\,s$^{-1}$. There is a hint of wings in these lines extending to about --20~km\,s$^{-1}$. The peak of the CO emission, especially in the higher $ J=3-2 $ transition is observed at more negative velocities. There is no detectable HCO$^+$(4--3), N$_2$H$^+$(3--2), CH$_3$CN and continuum emission (for N$_2$H$^+$ and  CH$_3$CN we use the data from Paper~I). 

\begin{figure}
\includegraphics[angle=-90,width=\columnwidth]{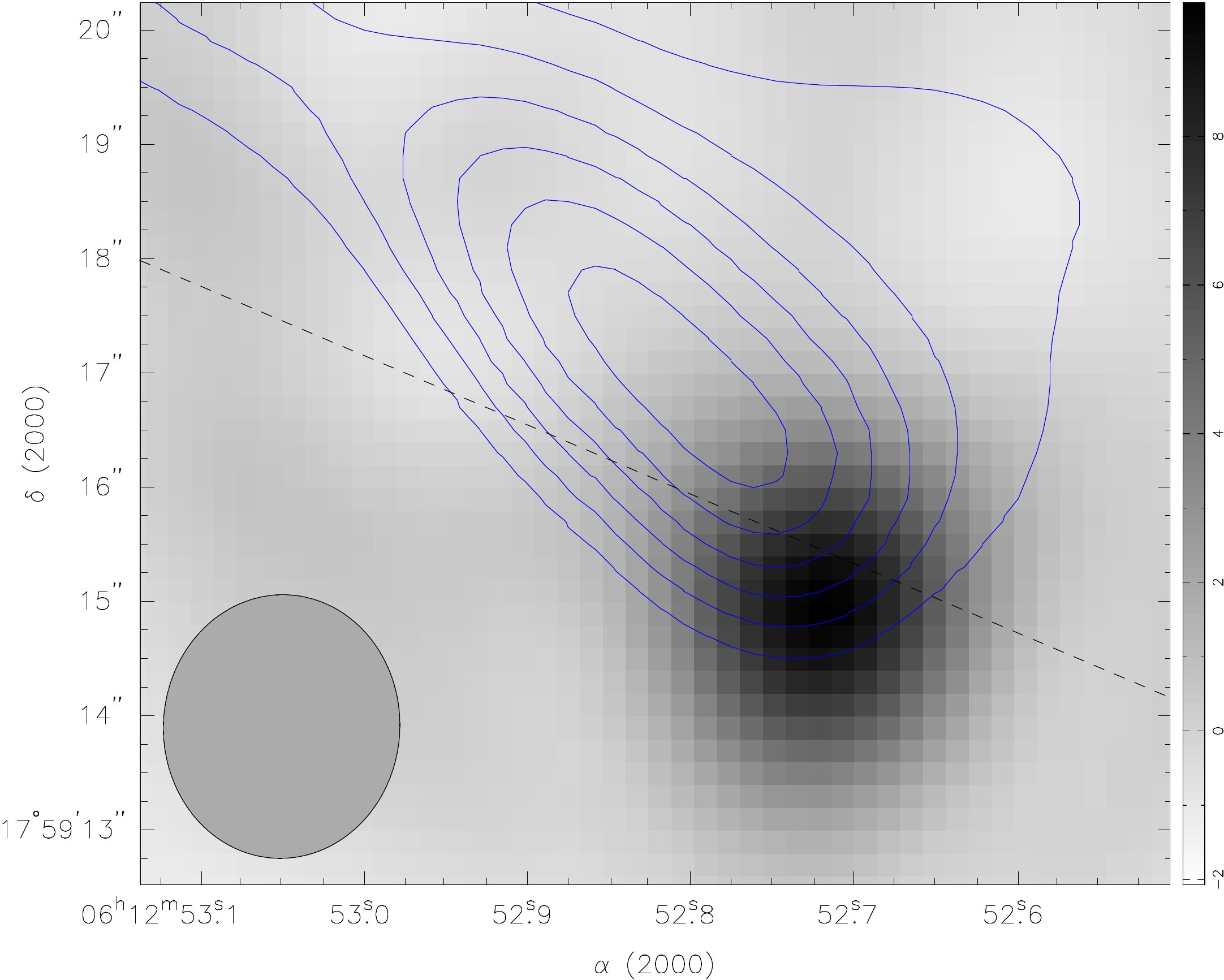}
\caption{The map of CO(3--2) emission in the velocity range --19...+1~km\,s$^{-1}$ (contours) overlaid on image of CS(7--6) emission in the velocity range --13...+1~km\,s$^{-1}$ (grey scale). The intensity units for the image are Jy\,beam$^{-1}$\,km\,s$^{-1}$. The contour levels are (3, 5, 7, 9, 11, 13)$\times 5$~Jy\,beam$^{-1}$\,km\,s$^{-1}$. {The SMA beam is shown in the lower left corner.}} 
\label{fig:hvc}
\end{figure}

The spatial distribution of the CS(7--6) and CO(3--2) emission integrated in the velocity ranges --13...+1~km\,s$^{-1}$ and --19...+1~km\,s$^{-1}$, respectively, is shown in Fig.~\ref{fig:hvc}. The HCN distribution is very similar to that of CS. The deconvolved size of the CS emitting clump is about $ 1{\farcs}1\times 0{\farcs}3 $ which corresponds to $ 1800\,\mathrm{AU} \times 500\,\mathrm{AU} $. It is apparently located at the head of the stream observed in CO, almost exactly at the jet axis shown in Fig.~\ref{fig:CO-30m}. 

At the velocity of the CS and HCN emission peak, the brightness temperatures in the CO(3--2), CO(2--1) and HCN(4--3) lines are practically the same. The frequencies of the CO(3--2), HCN(4--3) and CS(7--6) lines are close to each other and beam parameters for them should be similar, too. However, the comparison with the CO(2--1) is {complicated by the} significantly different frequencies. 



The CO(3--2) line is most probably saturated. Then the HCN(4--3) line should be saturated, too. The CS line is weaker implying either a relatively lower optical depth (but close to unity anyway) or a smaller size of the emission region. 
A simple modeling using e.g. RADEX \citep{vdTak07} shows that the column densities of HCN and CS required to explain such optical depths are $ \ga 10^{15} $~cm$^{-2}$. The relative abundances of these molecules are $ \le 10^{-8} $. Therefore the total gas column density in the clump is  $ \ga 10^{23} $~cm$^{-2}$. Then, using the size estimated above we obtain the gas density $ n \ga 3\times 10^{6} $~cm$^{-3}$. This estimate is consistent with the observations of the HCN and CS lines which require at least such densities for excitation.

The large HCN and CS line widths indicate that the clump is gravitationally unbound. The virial mass of this clump estimated in the usual way \citep[e.g.][]{Zin94} is $ M_\mathrm{vir} \sim 30 $~M$_\odot$. At the same time, {the non-detection} of dust emission implies an upper limit for mass orders of magnitudes lower. Therefore, the clump represents a transient entity. 

It is worth mentioning {the non-detection of} HCO$^+$ and N$_2$H$^+$ emission. Both molecules can be destroyed by dissociative recombination \citep[e.g.][]{Zin09}. Therefore their absence can indicate an enhanced ionization in the dense clump.


\subsection{Ionized gas}
In Fig.~\ref{fig:co32-outflow} we plot the map of the 15~GHz continuum emission near the SMA1/SMA2 clumps (from the VLA archival data, the angular resolution is about 4 arcsec) which shows the distribution of the ionized gas in this area. This ionized component seems to be associated with the jet traced in the \Fetwo\ emission. The GMRT map at 1280~MHz looks similar but there is a significant positional uncertainty in the GMRT data (Paper~I). Properties of the continuum source were estimated in Paper~I and by \citet{Ojha11}. The emission measure is $EM \sim (1-2.5)\times 10^7 $~pc\,cm$^{-6}$. Taking into account the observed size of the continuum source, the electron density is $ n \sim 3\times 10^4 $~cm$^{-3}$.

\section{Surroundings} 

The surroundings of the SMA1/SMA2 clumps and high velocity outflow are traced in several molecular lines. One of the most informative is the N$_2$H$^+$(3--2) transition observed with both SMA and IRAM-30m telescope. The combined map of the N$_2$H$^+$(3--2) emission is presented in Fig.~\ref{fig:CO-30m}. It shows an absence of N$_2$H$^+$ in the hot core, in accordance with our previous findings (Paper~I). The overall morphology of the N$_2$H$^+$ emission {suggests that it originates in} an envelope around the central cores and the outflow lobes. 

\begin{figure*}
\begin{minipage}{0.49\textwidth}
\centering
\includegraphics[angle=-90,width=\columnwidth]{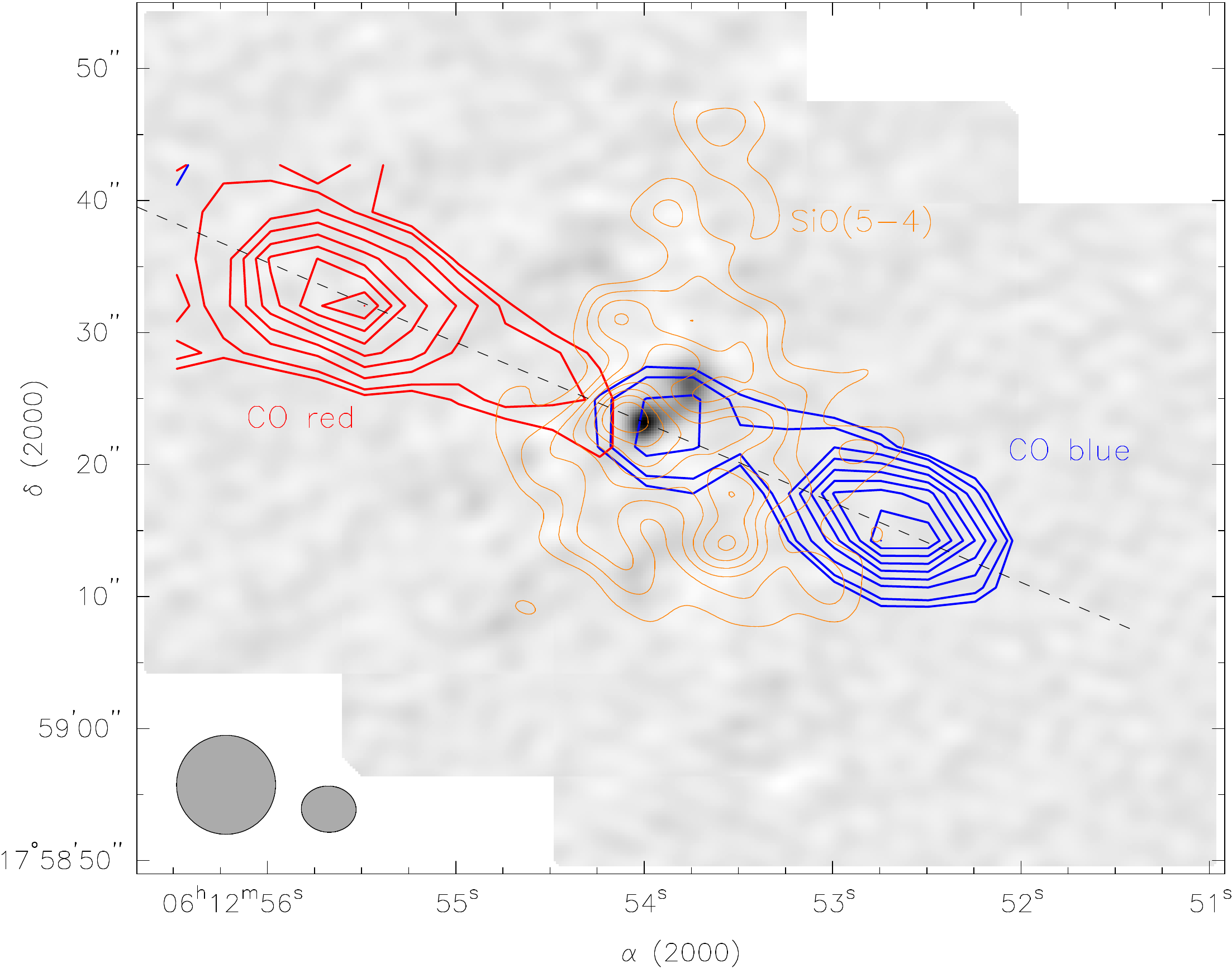}
\end{minipage}
\hfill
\begin{minipage}{0.49\textwidth}
\centering
\includegraphics[angle=-90,width=\columnwidth]{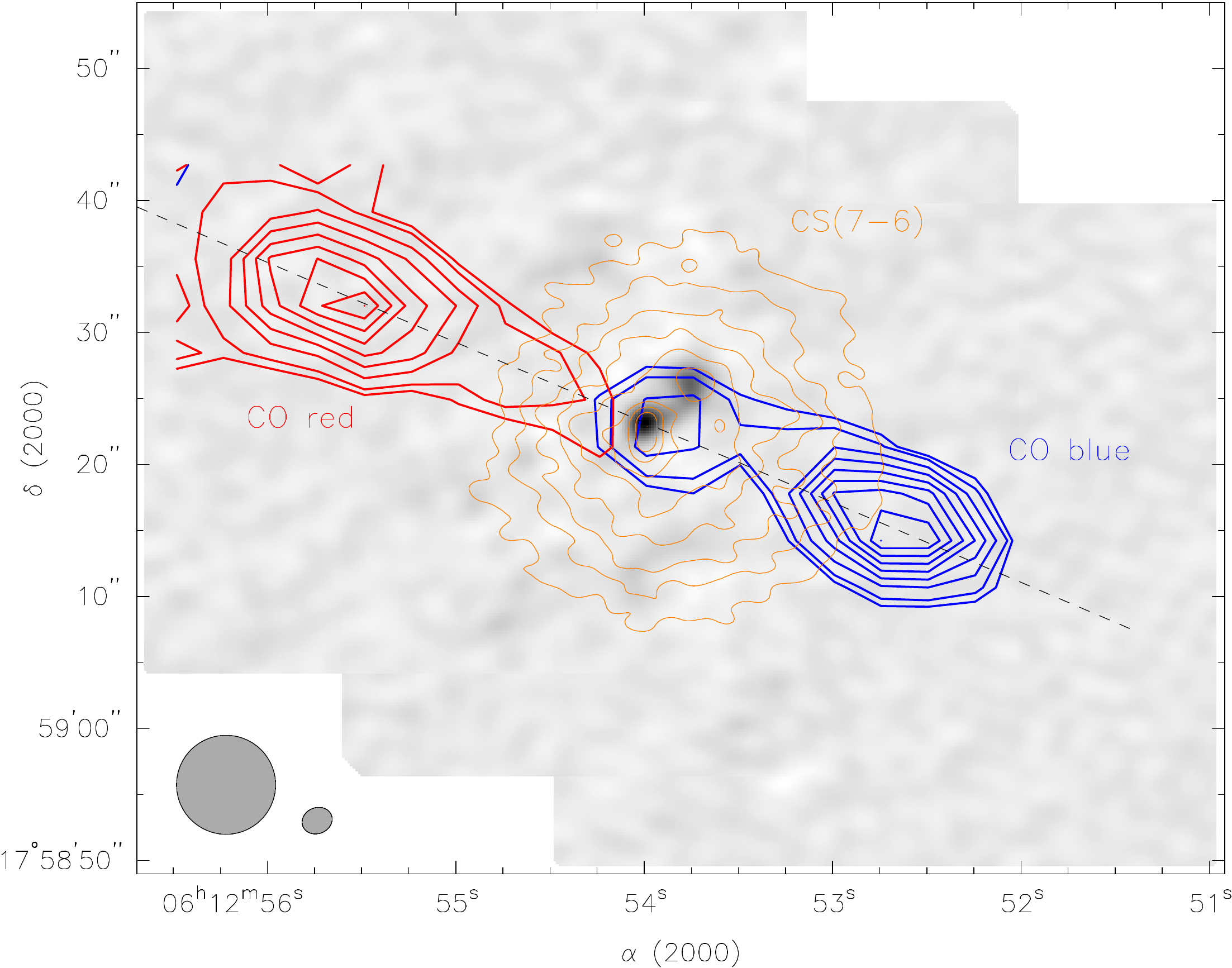}
\end{minipage}
\caption{Left panel: the maps of the SiO(5--4) (left panel) and CS(7--6) (right panel) integrated line emission obtained by combining the SMA and 30m data (orange contours) overlaid on the continuum image at 0.8~mm. The maps of the CO(3--2) high velocity emission as observed with the IRAM 30m telescope (blue and red thick contours) from Fig.~\ref{fig:CO-30m} are also shown. {The beams for the IRAM-30m telescope and for the combined maps are shown in the lower left corner of each panel.}}
\label{fig:SiO+CS}
\end{figure*}

The distribution of the SiO(5--4) emission is different (Fig.~\ref{fig:SiO+CS}). It peaks near the SMA1 core. There is also a feature in the area of the blue outflow lobe which may be associated with the N$_2$H$^+$ emission. The CS(7--6) distribution (right panel in Fig.~\ref{fig:SiO+CS}) seems to be the most uniform one. It shows a rather smooth, almost spherical halo around the SMA1 and SMA2 cores. There is no sign of the outflow influence on the CS distribution. One may suspect that the optical depth in the CS line is too high and we see only the outer regions of the core. However the SMA1 and SMA2 clumps are well resolved in this map which makes such explanation less probable. Then, the total gas distribution is apparently not significantly affected, too.

\section{Discussion} \label{sec:disc}
The main goal of this study is the characterization of the outflow and probable accretion disk associated with the massive YSO in the S255IR clump. The data presented in the previous sections shed new light on this system.  

Concerning the outflow, one of the main findings is that the SMA interferometric data alone give a rather distorted picture. They hint at two highly collimated parallel outflows with different centers of origin. However, this impression is apparently caused by a significant flux loss in the interferometric measurements. The single-dish CO observations clearly show a single less collimated outflow originating at the SMA1 core. The CO emission retrieved by the SMA near the driving source originates apparently from the northern wall of the outflow cavity. This means that this part of the wall is rather thin and bright. The absence of a noticeable CO emission at the opposite side of the wall implies a more diffuse distribution of the emission here, probably due to the density structure of the surrounding medium.

The question {arises of} how common is this effect in the interferometric studies of outflows. We can easily imagine a situation when an interferometric image will show multiple outflows from a single driving source, while in fact there is a single wide-angle outflow.

{The observed P--V diagrams for the outflow (Figs.~\ref{fig:CO-pv},\ref{fig:HCO-pv}) most probably indicate periodic ejections from the driving source. Two events can be traced in the data.}
They apparently created jet knots at different distances from the central star. The older one is responsible for the extended CO outflow. The peaks of the CO emission coincide with bright H$_2$ emission spots \citep{Wang11} which probably indicate bow shocks at the heads of the jets. The next ejection event created other bow shocks which are seen in particular in the \Fetwo\ emission (Fig.~\ref{fig:co32-outflow}). This later ejection entrains dense molecular gas observed in the wings of the HCO$^+$, HCN and CS lines. It is also traced in the CO emission as can be seen in the CO position-velocity diagram (Figs.~\ref{fig:CO-pv},\ref{fig:HCO-pv}). Another manifestation of this activity is apparently the dense high velocity clump (Sect.~\ref{sec:hvc}), which most probably represents dense gas at the head of bow shock. All the jet knots lay practically on {a straight} line. Therefore the orientation of the jet does not change with time significantly. The age of the first, extended outflow was estimated of about 7000 years by \citet{Wang11}. Assuming the same ejection velocities for the two events and comparing distances of the bow shocks from the driving source we can conclude that the second ejection happened about 1000 years ago. Therefore the time interval between the ejection events is about 6000 years (the age estimates were done under the usual assumption of an inclination angle for the jet of 45$^\circ$).  

Several entrainment mechanisms are usually discussed for molecular outflows \citep[e.g.][]{Arce07,Frank14}. In our case the most probable one is the jet bow-shock model \citep[e.g.][]{Raga93}. {The main argument in favor of this model is the obvious presence of several bow shocks (traced in the H$_2$ and \Fetwo\ emission) clearly associated with the high velocity molecular gas.} This model is also consistent with the observed outflow morphology and kinematics. 
The shape of the P--V diagram differs from the frequently observed ``Hubble law'' and shows a range of velocities at the largest distances from the driving source. {A similar shape was observed in some other outflows which are apparently driven by bow shocks and is expected in the theoretical models \citep[e.g.][]{Lee00}.}

The jet is apparently launched from the accretion disk around the central massive young star. As mentioned above, there is a chain of water masers along the jet. For several water maser spots shown in Fig.~\ref{fig:sma1_ch3oh-vel}, proper motions have been measured \citep{Goddi07}. The velocities (from $\sim 10$~km\,s$^{-1}$ to $\sim 30$~km\,s$^{-1}$ {at the distance of 1.6~kpc}) are mostly perpendicular to the jet axis with outward components in some cases. {The maser velocity pattern is consistent with} clockwise rotation around the core center. 
Velocities of the masers closest to the center imply dynamical mass of the order of 20~M$_\odot$. The mass of the central star is estimated to be 27~M$_\odot$ \citep{Ojha11} assuming the distance of 2.5~kpc. At the distance of 1.6~kpc the mass of the star would be $23.5 \pm 3.6$~M$_\odot$. {This estimate is based on the unpublished value of the star luminosity of $(6.8\pm 2.8)\times 10^4 $~L$_\odot$ (for the distance of 1.6~kpc). Now we reconsidered the star SED by including data at millimeter wavelengths from \citet{Zin09}. This results in the luminosity of $(3.5\pm 0.3)\times 10^4 $~L$_\odot$ and mass of $20 \pm 2$~M$_\odot$.} It is worth mentioning that \citet{Pashchenko03} estimated the mass of the central star to be (6--7)~M$_\odot$ from their analysis of water masers variability. However, they {employed a systemic} velocity of 8.3~km\,s$^{-1}$ which contradicts our findings which clearly show {the systemic} velocity to be about 4.8~km\,s$^{-1}$. {In the following we will take the mass of 20~M$_\odot$ as the most probable value.} This is sufficient to explain velocities of the masers closest to the star. However maser spots located at larger distances from the center have similar velocities and imply a much larger central mass of the order of 100~M$_\odot$. This value is inconsistent with all other estimates. The corresponding mean density would be unrealistically high. This means that we probably need another explanation for the maser velocities. {Rather speculative, it can be jet rotation. Rapid rotation of protostellar jets is expected in theoretical models \citep[e.g.][]{Pudritz07} and was indeed observed in some cases \citep{Bacciotti02}.}

The {water masers are thought to be} excited by interaction of the jet with the surrounding medium. The fact that they are moving in one direction on each side of the central star implies some asymmetry in the system. 
There may be some misalignment between the jet and the rotation axis of the material in the outer parts of the clump {(apparently, in the axially-symmetric case the observed maser movements would be symmetric relative the axis)}. It can be that in the inner part the orientation of the disk plane is different. However, the difference cannot be large. We see also other indications of the jet interaction with the surrounding medium, in particular, increased line widths along the jet. 



Fig.~\ref{fig:sma1_ch3oh-vel} shows that the water maser condensations are shifted from the jet axis in direction of their proper motion. This projected shift is of the order of 200~AU. At the velocities mentioned above, the maser condensations would travel this distance in about 30--100~years. From this consideration it looks probable that the event created these masers happened quite recently.

The disk is apparently strongly fragmented as follows from the derived beam filling factor ($ \sim 0.2) $ for {CH$_3$CN and CH$_3$OH}. The mean gas density in the disk is about $6\times 10^8$~cm$^{-3}$. The density of the fragments should be much higher. Observations of vibrationally excited HCN indicate densities $ > 10^{10} $~cm$^{-3}$ which is consistent with this picture.

The derived mass of the hot gas ($\sim 0.3$~M$_\odot$) is by an order of magnitude insufficient to gravitationally bound the core. The virial mass estimated in the usual way \citep[e.g.][]{Zin94} from the line width is about 7~M$_\odot$. 
There can be also some amount of cold material in the disk, as follows from the DCN observations. However, the total mass of the disk is apparently much lower than mass of the central star which is estimated of about 20~M$_\odot$ {(see above)}. Since the disk mass is significantly lower than the star mass, the disk is probably not self-gravitating and we may expect Keplerian rotation. Unfortunately the achieved angular resolution is not sufficient {to measure reliably the disk rotation curve}.

For a central mass of 20~M$_\odot$ the inclination angle of the disk (derived from comparison of the observed rotation line-of-sight velocity with the expected velocity for this mass) should be small, about 25$^\circ$ (Fig.~\ref{fig:sma1_ch3oh-pv}), i.e. the disk is seen almost face-on. This is consistent with our maps which do not show a significant elongation of this object. {On the other hand this conclusion looks somewhat suspicious because of the well-collimated appearance of the molecular outflow. Observations at a higher angular resolution are needed to clarify the structure and kinematics of this object.}

{The temperature of the hot gas derived from the CH$_3$CN, CH$_3$OH and SO$_2$ observations is about 130--180~K. A similar estimate for the temperature of the hot component was obtained by us many years ago from the analysis of the IRAS data \citep{Zin90}. The HNCO rotational temperature is somewhat higher, about 320~K. However, as mentioned above, this value probably represents an upper limit to the excitation temperature. In addition, the HNCO excitation can be influenced by the FIR radiation.} The HNCO excitation in massive cores was discussed by \citet{Zin00}. That analysis shows that in the present case collisional excitation can be effective. The critical density for excitation of the $ K_{-1} = 3 $ ladder is of the order of $ 10^{10} $~cm$^{-3}$. Such density is quite possible in the SMA1 core as follows from the estimates above.
{However the radiative excitation by FIR emission via the $b$-type transitions cannot be excluded, too. With the column density of $ N(\mathrm{H_2}) \sim 3\times 10^{24} $~cm$^{-2}$ the dust will be optically thick at these frequencies creating a sufficiently strong radiation field. The HNCO excitation especially in the higher $ K_{-1} $ ladders may reflect the effective temperature of this field.}

The highest temperature of the observed molecular material is about 300~K. {Taking the source} luminosity to be about $ 3.5\times 10^4 $~L$_\odot$ (see above) {this value} corresponds to the dust equilibrium temperature at the distance of 200--300~AU from the star. Since we do not see warmer molecular gas, {no dust should be present} within this radius in case of no shielding. However this {radius} is comparable to the observed size of the core and we do not see any central hole in the molecular distribution. {It is then possible that} shielded molecular clumps are {distributed} at smaller radii.  

In this respect an interesting feature of our observations is the apparently relatively cold and rather massive clump with a strong DCN emission. It may be gravitationally bound and the mass of the clump is sufficiently high to consider it as a possible low-mass protostar.

Concerning the surroundings of the SMA1 clump, we see that the outflow strongly affects the chemical composition of the medium. The N$_2$H$^+$ molecules are destroyed along the outflow. The SiO distribution seems to be also significantly affected. At the same time there is no noticeable influence on the CS distribution. The density structure is probably not affected, too. As mentioned above (Sect.~\ref{sec:sma2},\ref{sec:sma4}) the nearby clumps are probably influenced by shocks which are apparently associated with this outflow.

\section{Conclusions}
We presented the results of our observations of the S255IR area with the SMA at 1.3~mm in the very extended configuration and at 0.8~mm in the compact configuration as well as with the IRAM-30m at 0.8~mm. The best achieved angular resolution is about 0.4 arcsec. The dust continuum emission and several tens of molecular spectral lines are observed. The majority of the lines is detected only towards the S255IR-SMA1 clump.
In summary, our main findings are the following:

1. The S255IR-SMA1 clump represents apparently a rotating structure (probably a disk) around the young massive star. The achieved angular resolution is still insufficient {to establish the character (Keplerian or non-Keplerian)} of the rotation. The temperature of the molecular gas reaches 130--180~K. The size of the clump is about 500~AU. It is apparently strongly fragmented as follows from the derived small ($ \sim 0.2$) beam filling factors  for various molecules. The mean gas density is about $6\times 10^8$~cm$^{-3}$. The density of the fragments should be much higher which is confirmed by observations of HNCO and vibrationally excited HCN. The mass of the hot gas is $\sim 0.3$~M$_\odot$ and the total mass of the clump is significantly lower than the mass of the central star (about 20~M$_\odot$). The inclination angle of the disk (derived from comparison of the observed rotation line-of-sight velocity with the expected rotation velocity for this mass) should be small, about 25$^\circ$.

2. We detected a strong DCN(3--2) emission near the center of the SMA1 clump. Most probably it indicates the presence of a rather large amount ($\ga 1$~M$_\odot$) of cold ($ \le 80 $~K) material. This cold clump can be gravitationally bound.

3. High velocity emission is observed in the CO line as well as in lines of high density tracers HCN, HCO$^+$, CS and other molecules. The CO outflow is much more extended than that observed in the lines of high density tracers. Its morphology obtained from combination of the SMA and IRAM-30m data is significantly different from that derived from the SMA data alone. The CO emission detected with the SMA traces only one boundary of the outflow and leads to a rather distorted picture of the outflow structure. 
The velocity of the CO outflow reaches $ \sim 60 $~km\,s$^{-1}$. 

4. The outflow is most probably driven by the jet bow shock mechanism. The available data indicate at least two major ejection events with a time interval of several thousand years between them. The direction of the ejections does not change with time.
The high velocity emission in the lines of high density tracers is associated with the peaks of the \Fetwo\ emission related to the bow shocks caused by rather recent ejections from the SMA1. 

We detected a dense high velocity clump associated apparently with one of the bow shocks. It shows a strong emission in the HCN(4--3) and CS(7--6) lines. At the same time there is no detectable HCO$^+$(4--3) or N$_2$H$^+$(3--2) emission which can be probably explained by enhanced ionization.

5. The proper motions of the water masers excited along the jet imply some misalignment of the jet with the rotation axis of the material in the outer parts of the clump. It can be that the orientation of the disk in the inner and outer parts is somewhat different. However, the difference cannot be large.

6. The outflow strongly affects the chemical composition of the surrounding medium. The N$_2$H$^+$ molecules are destroyed along the outflow. The SiO distribution seems to be also significantly affected. At the same time there is no sign of the outflow influence on the CS distribution. The total gas distribution is apparently not significantly affected, too.


\acknowledgments
This work was supported by the Russian Academy of Sciences (Research program No. 17 of the Department of Physical Sciences), Russian Foundation for Basic Research (RFBR), National Science Council (NSC) of Taiwan and Department of Science and Technology (DST) of the Government of India in frameworks of the research grants RFBR 08-02-92001-NSC, RFBR 13-02-92697-Ind, RFBR 15-02-06098, 15-52-45057, NSC 97-2923-M-001-004-MY3, DST-RFBR INT/RUS/RFBR/P-142. A.M.S. was supported by the Russian Science Foundation (grant number 15-12-10017). S.V.S. was supported by the Ministry of Education and Science of the Russian Federation (state task No. 3.1781.2014/K). Y.W. was supported by Swiss National Science Foundation, NSFC 11303097 and 11203081, China. The research is also partly supported by the grant within the agreement No. 02.В.49.21.0003 between The Ministry of Education and Science of the Russian Federation and Lobachevsky State University of Nizhni Novgorod and by Russian Education and Science Ministry Project 3.1252.2014/k. We are grateful to Elena Trofimova for the help with the line identification and to the anonymous referee for the detailed helpful comments.
The research has made use of the SIMBAD database,
operated by CDS, Strasbourg, France.

\bibliographystyle{apj}
\bibliography{apj-jour,s255}

\end{document}